\documentclass[prd,twocolumn,showpacs,preprintnumbers,aps,nofootinbib]{revtex4-1}

\usepackage{graphicx}% Include figure files
\usepackage{dcolumn}% Align table columns on decimal point
\usepackage{bm}% bold math
\usepackage{amsmath,amssymb,amsfonts}
\usepackage{latexsym}

\usepackage{color}
\def\nn    {\nonumber}

\begin{document}

%\preprint{APS/123-QED}

\title{\boldmath
Muon Flavor Violation in Two Higgs Doublet Model
 with Extra Yukawa Couplings
}

\author{Wei-Shu Hou and Girish Kumar}
\affiliation{
Department of Physics, National Taiwan University, Taipei 10617, Taiwan}
% \affiliation{ }
\bigskip

\date{\today}

\begin{abstract} 
The new round of experiments, MEG~II, COMET/Mu2e, and Mu3e, would soon 
start to push the $\mu \to e\gamma$, $\mu N \to eN$ conversion, and $\mu \to 3e$ frontier, 
while Belle~II would probe $\tau \to \mu\gamma$ and $\tau \to 3\mu$.
In the general two Higgs doublet model with extra Yukawa couplings,
we show that all these processes probe the lepton flavor violating (LFV)
dipole transition that arises from the two loop mechanism, 
with scalar-induced contact terms subdominant.
This is because existing data suggest the extra Yukawa couplings
$\rho_{\mu e},\, \rho_{ee} \lesssim \lambda_e$, 
while $\rho_{\tau\mu},\, \rho_{\tau\tau} \lesssim \lambda_\tau$
and $\rho_{tt} \lesssim \lambda_t$, with $\lambda_i$ 
the usual Yukawa coupling of the Standard Model (SM), where
$\rho_{\mu e}\rho_{tt}$ and $\rho_{\tau\mu}\rho_{tt}$ enter 
the $\mu e\gamma$ and $\tau\mu\gamma$ two loop amplitudes, respectively.
{With the $B_s \to \mu\mu$ decay rate basically consistent with 
SM} expectation, together with the $B_s$ mixing constraint, 
we show that $B_s \to \tau\tau$ would also be consistent with SM,
while $B_s \to \tau\mu$ and $B \to K\tau\mu$ decays would
be out of reach of projected sensitivities,
in strong contrast with some models motivated by the B anomalies.
\end{abstract}

\maketitle

%-----------------------------------------------------------------------------------------------------------------------------------
%	Introduction
%-----------------------------------------------------------------------------------------------------------------------------------

\section{Introduction}

The study of muon properties is practically 
the oldest subject of particle physics, but 
remains at the forefront of current research.
The MEG bound~\cite{TheMEG:2016wtm} on 
muon flavor violating ($\mu$FV) $\mu \to e\gamma$ decay rate
at 90\% C.L. is
\begin{align}
{\cal B}(\mu \to e\gamma) < 4.2 \times 10^{-13}, \ \ ({\rm MEG}, 2016),
\label{eq:MEG16}
\end{align}
while a rather dated result of SINDRUM gives~\cite{Bellgardt:1987du}
\begin{align}
{\cal B}(\mu \to 3e) < 1.0 \times 10^{-12}, \ \ ({\rm SINDRUM, 1988}),
\label{eq:SINDRUM88}
\end{align}
for $\mu^+ \to e^+e^-e^+$ search.
A third type of $\mu$FV search studies 
$\mu \to e$ conversion on nuclei. Normalized to the muon capture rate,
SINDRUM~II finds~\cite{Bertl:2006up}
\begin{align}
R_{\mu e} < 7 \times 10^{-13}, \ \ ({\rm SINDRUM~II, 2006}),
\label{eq:SINDRUM06}
\end{align}
for $\mu \to e$ conversion on gold.

% planned $\mu\to e$ probes 

With schedules delayed by the current world pandemic,
MEG~II~\cite{Baldini:2018nnn}
will push the $\mu \to e\gamma$ bound down to 
$\sim 6 \times 10^{-14}$ with three years of data taking.
A new experiment to search for $\mu^+ \to e^+e^-e^+$, 
Mu3e~\cite{Blondel:2013ia}, plans to reach down to $5 \times 10^{-15}$ 
with three years of running and is limited mostly by the muon beam intensity.
Projected intensity improvements~\cite{Baldini:2018uhj} 
by up to 2 orders of magnitude seem feasible; hence, 
Mu3e can eventually reach down to $10^{-16}$ in sensitivity.
In contrast, to improve $\mu \to e\gamma$ sensitivity beyond MEG II,
innovations are needed for background suppression.

In terms of projected improvements, $\mu \to e$ conversion
i.e., $\mu N \to eN$ is perhaps the most promising.
SINDRUM~II operated at the limits of power consumption,
so new developments~\cite{DeeMe} are based on 
the idea~\cite{Dzhilkibaev:1989zb} of using special
solenoids for pion capture, muon transport, as well as detection,
which significantly improves muon intensity.
Phase I of COMET~\cite{Adamov:2018vin} 
aims for $R_{\mu e} < 7 \times 10^{-15}$,
eventually reaching down to $10^{-17}$ for phase II.
Similar to COMET phase II in design, Mu2e~\cite{Bartoszek:2014mya} 
aims at $2.6 \times 10^{-17}$ sensitivity.
Both experiments can be improved further.
For example, ongoing~\cite{Baldini:2018uhj} 
PRISM/PRIME~\cite{Kuno:2005mm} developments aim at 
bringing the limit eventually down to a staggering $10^{-19}$.
Although the primary objective for $\mu N \to eN$ is contact interactions, 
it also probes~\cite{deGouvea:2013zba} the dipole interaction
and can be in place to probe $\mu \to e\gamma$ if the associated backgrounds 
of the latter cannot be brought under control at high muon intensity.

{%\small
%\renewcommand{\arraystretch}{1.1}
%================================================
\begin{table*}[t!]
\begin{center}
%================================================
\begin{tabular}{|c|l|l|}
\hline
\ $\mu$FV process \ &
 \quad\quad\quad \ Current bound & 
 \quad\quad \ \ Future sensitivity   \\	
\hline
\hline
$\mu\to e \gamma$	&  
\ $4.2\times 10^{-13}$ (MEG~\cite{TheMEG:2016wtm}) 	&  
\ $6\times 10^{-14}$ (MEG II~\cite{Baldini:2018nnn})  	\\
$\mu \to 3 e$	&  
\ $1.0\times 10^{-12}$ (SINDRUM~\cite{Bellgardt:1987du}) \ 	& 
\ $\sim 10^{-15}{\rm -}10^{-16}$ (Mu3e~\cite{Blondel:2013ia})  	\\
$\mu N \to eN$ & 
\ \ $7 \times 10^{-13}$ (SINDRUM~II~\cite{Bertl:2006up}) \ & 
%\ \ $10^{-14}$ (DeeMe~\cite{Nguyen:2015vkk})    \\
% & & 
\ $\sim 10^{-15}{\rm -}10^{-17}$ (COMET~\cite{Adamov:2018vin})  \\
& & 
\ $3 \times 10^{-17}$-- \,\;\ (Mu2e~\cite{Bartoszek:2014mya}) \\
& & 
\ $\sim 10^{-18}{\rm -}10^{-19}$ (PRISM~\cite{Kuno:2005mm}) \  \\
\hline
\hline %\vskip0.05cm
$\tau \to \mu \gamma$	& 
\ $4.4\times 10^{-8}$ (BaBar~\cite{Aubert:2009ag})	 & 
\ ${\sim 10^{-9}}$ (Belle~II~\cite{Kou:2018nap})		\\
$\tau \to 3 \mu$	& 
\ $2.1\times 10^{-8}$ (Belle~\cite{Hayasaka:2010np})	 & 
\ $3.3 \times 10^{-10}$ (Belle~II~\cite{Kou:2018nap})		\\
\hline
\hline
\end{tabular}
\caption{Summary of current experimental bounds and 
future sensitivities of %charged lepton flavor violating (cLFV)
$\mu$FV processes.}
\end{center}
\label{tab:cLFV}
\end{table*}
}

The current bounds and projected sensitivities on 
$\mu$FV processes are summarized in Table~\ref{tab:cLFV}.
The impressive bounds for the muon reflect 
seven decades of studies.
We also list the corresponding processes for $\tau$,
i.e., $\tau \to \mu\gamma$ and $\tau \to 3\mu$, where the 
current bounds are from B factories~\cite{Aubert:2009ag,Hayasaka:2010np}, 
and expectations~\cite{Kou:2018nap} are for
Belle~II with 50~ab$^{-1}$ in the coming decade.
LHCb can~\cite{Bediaga:2018lhg} cross check the Belle~II 
result on $\tau \to 3\mu$ after upgrade~II, 
i.e., at the High Luminosity LHC (HL-LHC).
The heaviness of $\tau$ hence its later discovery, and 
smaller production cross section plus the difficulty in detection
underlie the weaker search limits.
However, its heavy mass and third generation nature offers
a different window on new physics, or equivalently,
beyond the Standard Model (BSM) physics.

% 1-2 vs 3-2 sectors of extra Yukawas 

We studied~\cite{Hou:2020tgl} the $\tau \to \mu\gamma$ decay 
previously in conjunction with $h \to \tau\mu$, where $h$ is 
the 125 GeV boson discovered in 2012~\cite{PDG}.
The context was the two Higgs doublet model (2HDM)
with extra Yukawa couplings, which was called the general 2HDM (g2HDM).
The $h$ boson picks up the extra $\rho_{\tau\mu}$ Yukawa coupling
from the $CP$-even exotic Higgs boson $H$ via $h$-$H$ mixing.
Given that this mixing angle, $c_\gamma$, is known to be small 
 (the alignment phenomenon~\cite{Hou:2017hiw}, or
 that $h$ so closely resembles the SM Higgs boson~\cite{PDG}),
only a weak constraint is placed on $\rho_{\tau\mu}$.
Together with the extra top Yukawa coupling $\rho_{tt}$, 
the $\rho_{\tau\mu}$ coupling induces $\tau \to \mu\gamma$ decay
via the two-loop mechanism~\cite{Chang:1993kw}.
Taking $\rho_{tt} \sim \lambda_{t} \simeq 1$,
the strength of the top Yukawa coupling of SM,
it was shown that Belle~II can probe the 
$\rho_{\tau\mu} \lesssim \lambda_{\tau} \simeq 0.010$ parameter space.

Taking $\rho_{tt}$ at ${\cal O}(\lambda_t)$ and 
$\rho_{\tau\mu} \lesssim \lambda_{\tau}$ together, 
they correspond to~\cite{Hou:2020tgl}
\begin{align}
 \rho_{3j}^f \lesssim \lambda_3^f, \ \ \ (j \neq 1),
\label{eq:rho3j}
\end{align}
with $ \rho_{31}^f \ll \lambda_3^f$ expected.
{As we will see, this relation does not hold for down-type quarks  
because of tight constraints from ($K$ and) $B$ meson physics.}
The  probe of $\rho_{tt}$ by $\tau \to \mu\gamma$
via the two-loop mechanism is quite significant, as $\rho_{tt}$ 
can drive~\cite{Fuyuto:2017ewj} electroweak baryogenesis (EWBG),
or the disappearance of antimatter in the very early Universe.
A backup mechanism~\cite{Fuyuto:2017ewj} is through 
$|\rho_{tc}| \sim \lambda_t$ [{i.e.,~saturating} Eq.~(\ref{eq:rho3j})]
in case $\rho_{tt}$ accidentally vanishes. 

In this paper, we show that the MEG~II search for $\mu \to e\gamma$ 
would continue to probe
\begin{align}
 \rho_{\mu e} \lesssim \lambda_e,
\label{eq:rhomue}
\end{align}
which echoes $|\rho_{ee}| \sim \lambda_e \cong 0.0000029$ 
that is suggested~\cite{Fuyuto:2019svr} by the recent 
ACME result~\cite{Andreev:2018ayy} on 
electron electric dipole moment (eEDM), where a correlation of 
$|\rho_{ee}/\rho_{tt}| \propto \lambda_e/\lambda_t$ is implied.
That is, the tiniest $CP$ violation on Earth seems linked with 
the baryon asymmetry of the Universe (BAU)!
The $\rho_{\mu e}$, $\rho_{ee}$ behavior suggest
\begin{align}
 \rho_{i1}^f \lesssim \lambda_1^f, \quad\; 
\label{eq:rhoi1}
\end{align}
which likely holds also for $i = 3$, and seems plausible for $f = u, d$.
Thus, the affinity of the 1-2 sector of extra Yukawa couplings 
may be with the first generation, while the affinity of the 3-2 sector 
may be with the third generation, which echo the mass-mixing hierarchy.
{That the $\rho^d$ matrix is close to diagonal is a mystery.}

% vs B-anomalies

If the ``septuagenarian''
 (``octogenarian'' if counting from date of discovery) muon
appear ``sanitized'', i.e., very much SM-like, as reflected in 
the weak strength of the extra Yukawa couplings mentioned,
one cannot but think of the ``$B$ anomalies" that have been 
in vogue for almost the past decade. 
For a brief summary---and {\it critique}---of these $B$ anomalies; 
see, e.g., the ``{\it HEP perspective and outlook}"
given by one of us in the summer 2018~\cite{Hou:2019dgh}; 
the situation about the $B$ anomalies has not changed by much since then.
Some of the suggested remedies of the $B$ anomalies, especially
the leptoquark (LQ) variant, relate to tree level effects, 
hence make a large impact in general.
In contrast, though also at tree level, the extra Yukawa couplings 
have hidden themselves so well for decades, via the relations 
such as Eqs.~(\ref{eq:rho3j}) and (\ref{eq:rhoi1}), 
{the near-diagonal $\rho^d$ matrix, 
{\it plus alignment}~\cite{Hou:2017hiw}.
A second purpose} of the present paper  is therefore 
to contrast the predictions of g2HDM vs the ``bold'', UV-complete models 
such as PS$^3$~\cite{Bordone:2017bld,Bordone:2018nbg,Cornella:2019hct}.
For this reason, we will extend the list of $\mu$FV processes
beyond Table~I to include various rare (semi)leptonic $B$ decays.

\begin{figure*}[t]
\center
\includegraphics[width=0.275 \textwidth]{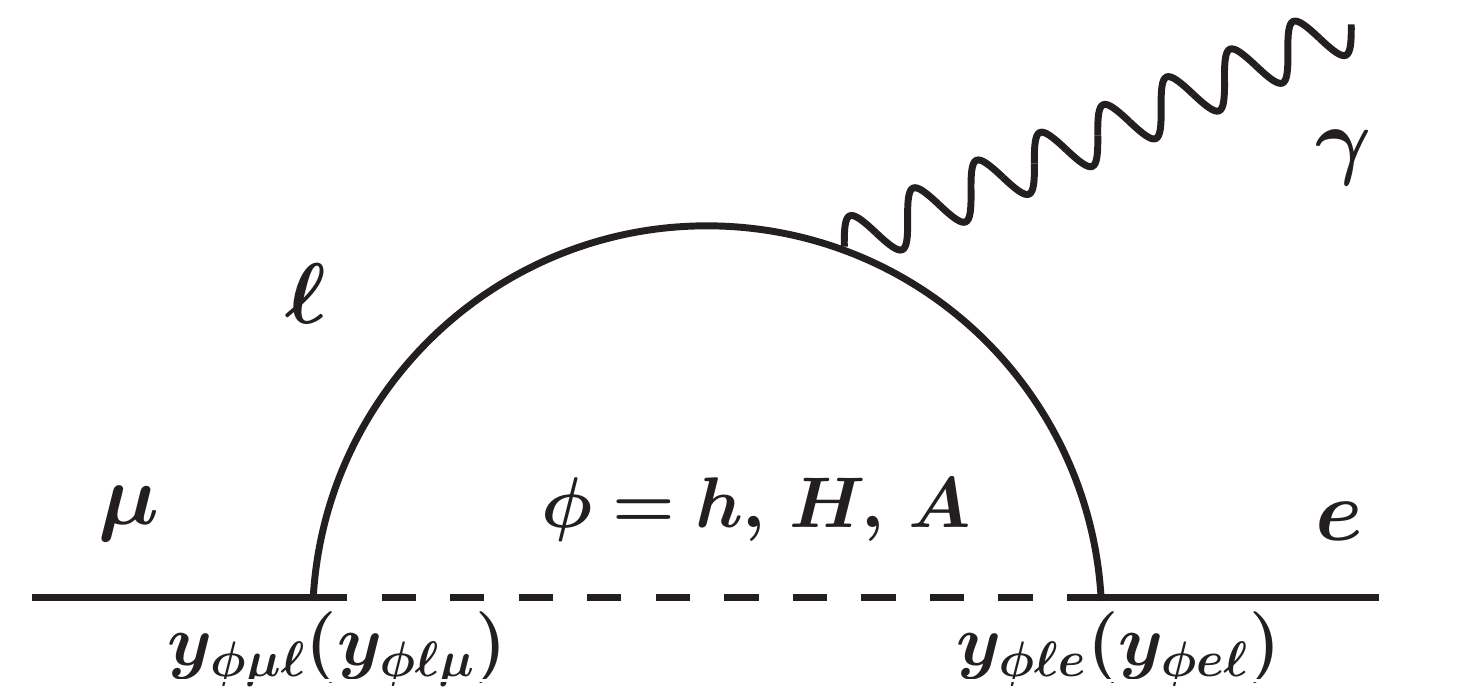}
\hskip0.35cm
\includegraphics[width=0.275 \textwidth]{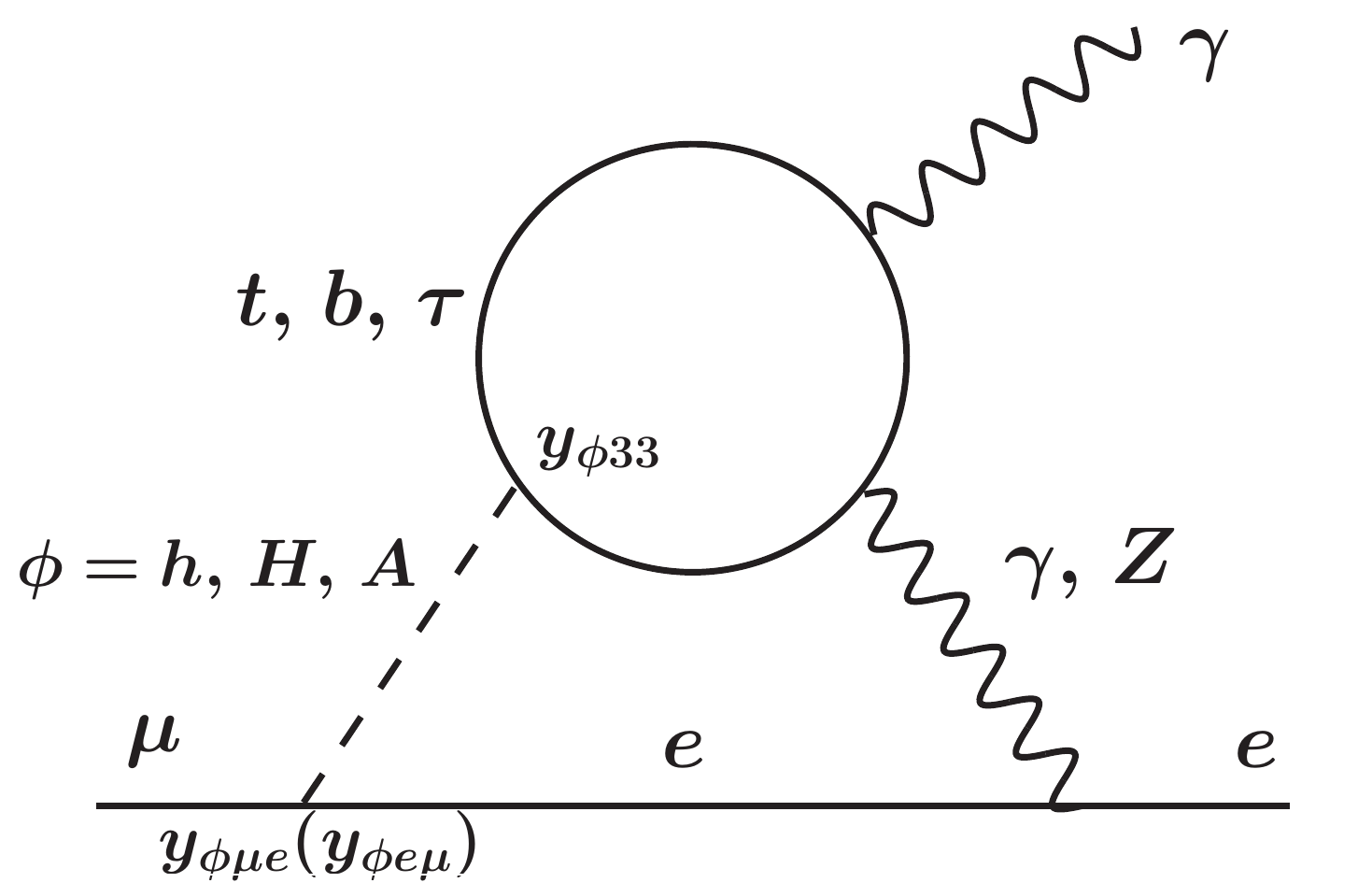}
\hskip0.35cm
\includegraphics[width=0.275 \textwidth]{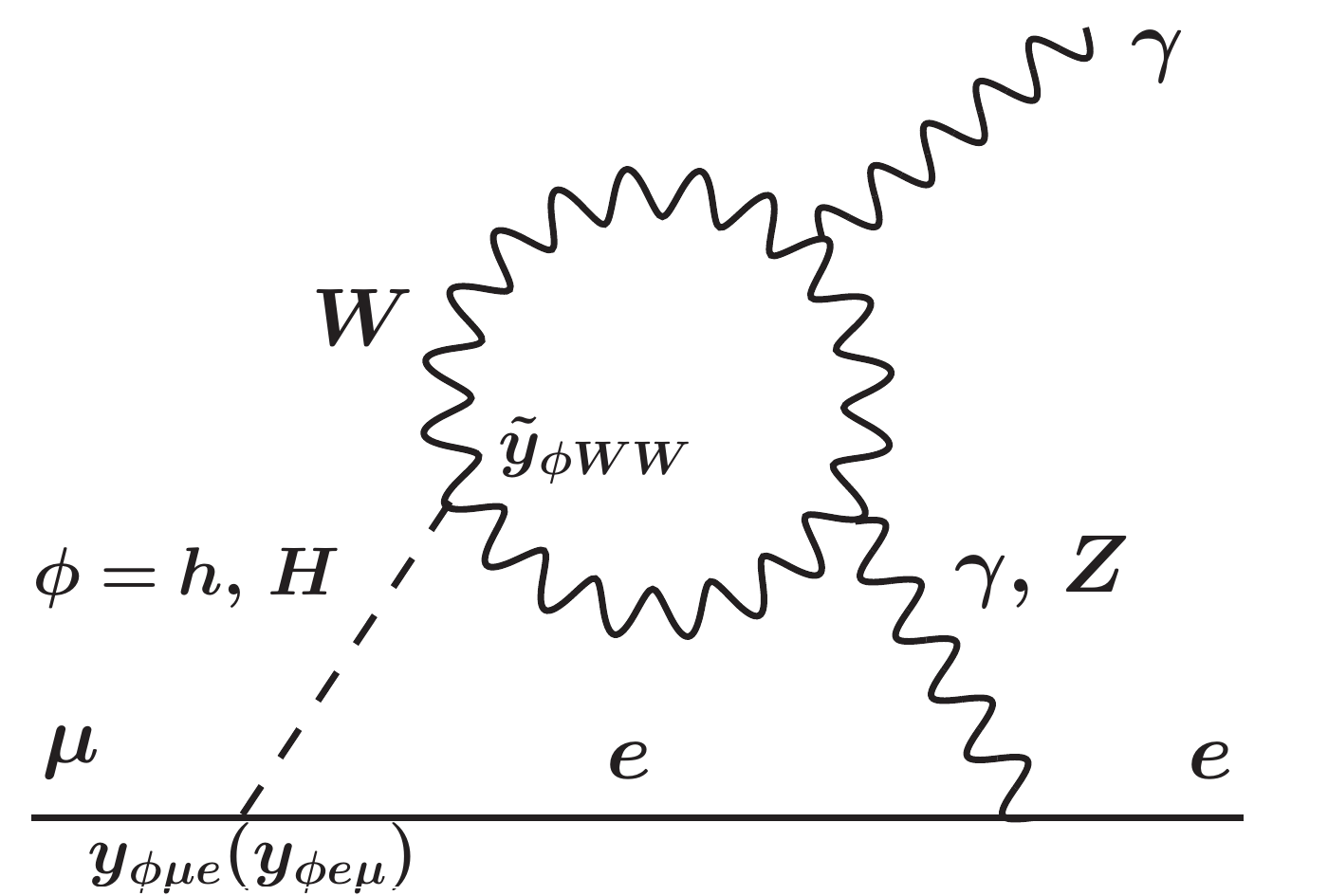}
\caption{
One-loop, two-loop fermion, and two-loop $W$ diagrams for $\mu\to e\gamma$.}
\label{fig:feyndiag}
\end{figure*}

The paper is organized as follows. 
In the next section, 
we discuss $\mu \to e\gamma$ in g2HDM,  
which is pretty much parallel to what we have done for 
$\tau\to \mu\gamma$~\cite{Hou:2020tgl}.
We show that the $\mu \to e\gamma$ process probes 
the  $\rho_{\mu e}\rho_{tt}$ product in g2HDM,
{as well as $c_\gamma \rho_{\mu e}$ where $c_\gamma$ is the $h$-$H$ mixing angle.}
In Sec.~III, we cover the $\mu \to 3e$ and $\mu N \to eN$ 
processes, as well as $\tau \to 3\mu$. 
{We show that the g2HDM effects are very suppressed at tree level  
and that all these processes eventually pick up the $\mu e\gamma$
or $\tau \mu \gamma$ dipole couplings.
}
In Sec.~IV, we contrast the pr{oje}ctions of g2HDM with the  
PS$^3$ model~\cite{Cornella:2019hct} motivated by the $B$ anomalies,
covering rare $B$ decays such as $B_q \to \tau\tau$, $\tau\mu$, 
$B \to K^{(*)}\tau\tau$, $K^{(*)}\tau\mu$,
 and {$\tau\to \mu\gamma$ as well}. %, $3\mu$
We also mention $B \to \mu\nu,\; \tau\nu$ decays,
where g2HDM could actually reveal~\cite{Hou:2019uxa} itself.
We briefly touch upon muon EDM and $g-2$, 
before offering our conclusion in Sec.~V.

\section{\boldmath The $\mu \to e\gamma$ Process}

MEG~II~\cite{Baldini:2018nnn} has a genuine
discovery potential in g2HDM with extra Yukawa couplings.

We have studied~\cite{Hou:2020tgl} $\tau \to \mu\gamma$ decay previously 
and showed that $\rho_{\tau\mu} \lesssim \lambda_\tau \simeq 0.010$ 
[part of Eq.~(\ref{eq:rho3j})] can be probed by Belle~II
as it pushes down to ${\cal O}(10^{-9})$~\cite{Kou:2018nap}.
The $\mu \to e\gamma$ process is the template for 
$\tau \to \mu\gamma$ decay, for which the two loop mechanism
 (see Fig.~\ref{fig:feyndiag}) of Ref.~\cite{Chang:1993kw} was {originally} 
written in g2HDM (called model III~\cite{Hou:1991un} at that time) 
that possesses extra Yukawa couplings.

{Our emphasis is on phenomenological discussion,}
so we take Ref.~\cite{Hou:2020tgl} as a template 
and do not recount details of the g2HDM here.
The formulas used in Ref.~\cite{Hou:2020tgl},
besides originating from Ref.~\cite{Chang:1993kw},
have also been checked against those of Ref.~\cite{Omura:2015xcg},
although one should use caution with this reference,
as it was written in a time when there was a hint for $h \to \tau\mu$ from CMS, 
which has subsequently disappeared~\cite{PDG}. 
{What should be emphasized is that, in g2HDM, the exotic Higgs bosons
$H$, $A$ ($CP$-odd), and $H^+$ would {\it naturally} populate
the 300--600\;GeV range but which we have surprisingly little knowledge of. For example, $H, A$ could be searched for in $t\bar{c}\, (\bar t c)$ \cite{Altunkaynak:2015twa} and $\tau\mu$ \cite{Hou:2019grj, Primulando:2016eod, Primulando:2019ydt} final states.}

\begin{figure}[b]
\centering
\includegraphics[angle=0,width=8.65cm]{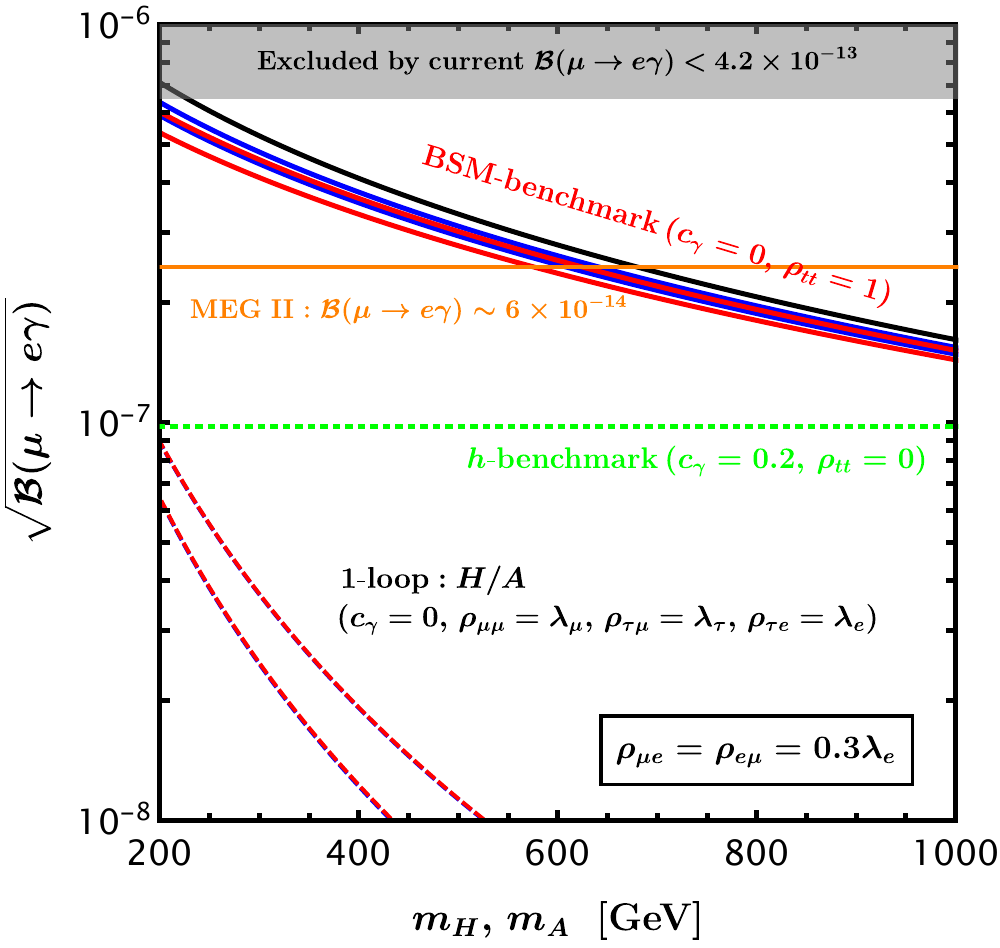}
\caption{
Comparison of benchmark scenarios for $\mu\to e\gamma$ 
as function of scalar masses. 
For one loop 
{red dashed curves, lower (upper) curve is   
for $m_A = m_H + 100\; (200)$~GeV, 
and flipping $H \leftrightarrow A$ is not much different.
For the two-loop BSM benchmark,
black curve is for degenerate $m_H = m_A$, 
red (blue) curves show variation in $m_H\, (m_A)$ 
%the other scalar mass 
with $m_A\, (m_H)$ heavier by} 100, 200~GeV,
where satisfying the MEG bound~\cite{TheMEG:2016wtm} 
at the low 200~GeV fixes $\rho_{\mu e} = \rho_{e\mu} \simeq 0.3 \lambda_e$.
Holding this value fixed, the two-loop $h$ benchmark is the green dashed horizontal line,
which lies below the MEG~II~\cite{Baldini:2018nnn} sensitivity.
See text for further discussion.
}
\label{benchmarks}
\end{figure}
%

% 1-loop: blunt instrument for rho_mutau * rho_taue

In g2HDM, flavor changing neutral Higgs (FCNH) couplings
are controlled~\cite{Hou:1991un} by the mass-mixing hierarchy; 
hence, the one loop diagram, Fig.~\ref{fig:feyndiag}(left),
is expected to be highly suppressed~\cite{Chang:1993kw}
by multiple chirality flips.
Using the one loop formula of Ref.~\cite{Hou:2020tgl} 
with a simple change of indices, we assume $\rho_{\mu\mu}\rho_{\mu e}$ from 
an intermediate muon in the loop is negligible compared with 
$\rho_{\tau\mu}^*\rho_{\tau e}$ from an intermediate $\tau$, 
which is even more so the case for an intermediate $e$.
We illustrate this ``one loop benchmark'' in Fig.~\ref{benchmarks}
for %
{$\rho_{\tau\mu} = \rho_{\mu\tau} = \lambda_\tau$ 
and $\rho_{\tau e} = \lambda_e$, and for $m_A = m_H + 200$\,GeV}
 (or with $H \leftrightarrow A$ interchanged).
The effect by itself is out of reach for any time to come,
unless $A$, $H$ are very light.
In fact, for $m_A = m_H \in (300,\, 500)$\,GeV, 
due to a cancellation mechanism, the MEG or the future MEG~II 
bounds  would allow $\rho_{\tau\mu}\rho_{\tau e}$ at 
${\cal O}(10^4)$ times larger than $\lambda_e\lambda_\tau$, 
which is very accommodating.
For nondegenerate $m_H = 300$\,GeV, $m_A = 500$\,GeV,
we find $\rho_{\tau\mu}\rho_{\tau e}/ \lambda_e\lambda_\tau \lesssim 17$ 
by MEG can be improved  to 6.6 with MEG~II, 
with the results similar for flipping $H \leftrightarrow A$.

% 2-loop: just scaling from tau -> mugamma
% 

It is the two loop mechanism~\cite{Chang:1993kw} that is 
of interest for g2HDM, where the $\rho_{\mu e}$ coupling induces 
$\mu \to e\gamma$ decay by inserting the $\phi \to \gamma V^*$ vertex 
[$\phi = h, H, A$; see Fig.~\ref{fig:feyndiag}(center) and 1(right)] 
related to the $h \to \gamma\gamma$ process, with $V = Z$ subdominant. 
Following Ref.~\cite{Hou:2020tgl} for $\tau\to\mu\gamma$, 
we define two BSM benchmarks for illustrating two loop effects.
Taking the extra top Yukawa coupling $\rho_{tt} \simeq 1$ 
while setting $c_\gamma = 0$,
one maximizes the $H$, $A$ effect but decouples the $h$ boson.
This ``BSM benchmark'' is illustrated in Fig.~\ref{benchmarks},
where $\rho_{\mu e} = \rho_{e\mu} \simeq 0.3\lambda_e$ is taken
to satisfy the current MEG bound of  Eq.~(\ref{eq:MEG16})
at $m_H$ or $m_A = 200$\,GeV.
The MEG~II experiment will continue to probe $\rho_{\mu e}$
down to lower values.

A second benchmark illustrates the effect of the SM-like $h$ boson,
where we take $\rho_{tt} = 0$ to decouple the exotic $H$, $A$ scalars,
but take $c_\gamma = 0.2$ as a large value that may still be allowed.
This ``$h$ benchmark'' is also plotted in Fig.~\ref{benchmarks},
giving ${\cal B}(\mu \to e\gamma) \simeq 10^{-14}$
for $\rho_{\mu e} = \rho_{e\mu} \simeq 0.3\lambda_e$,
which appears out of reach for MEG~II.
Depending on whether $c_\gamma$ is smaller or larger than 0.2,
the rate would drop further or become larger, although a 
$c_\gamma$ value larger than 0.2 may not be plausible.
But the rate scales only with the product of 
$c_\gamma^2\rho_{\mu e}^2$, and if $\rho_{tt}$ truly vanishes,
a $\rho_{\mu e}$ value larger than $0.3 \lambda_e$ is allowed.

We note that, unlike the $\tau \to \mu\gamma$ case where 
$h \to \tau\mu$~\cite{PDG} provides 
{a constraint~\cite{Hou:2020tgl}
on $c_\gamma\rho_{\tau\mu}$, no realistic constraint 
on $c_\gamma\rho_{\mu e}$} can be extracted from 
$h \to \mu e$ search~\cite{PDG} for our purpose, 
as $\mu \to e\gamma$ already constrains $\rho_{\mu e}$ to be so small.
On the other hand, the value of $\rho_{tt}$ is not known at present,
except that  any {\it finite} value may suffice~\cite{Fuyuto:2017ewj} 
for EWBG. For instance, in trying to account for the strong 
bound on electron EDM by ACME~\cite{Andreev:2018ayy}, 
the smaller $|\rho_{tt}| \simeq 0.1$ was chosen in Ref.~\cite{Fuyuto:2019svr} 
to ease the tension. While $\rho_{tt}$ at ${\cal O}(1)$ is not strictly ruled out,
we stress that $\mu \to e\gamma$ probes the $\rho_{\mu e}\rho_{tt}$ 
product; hence, we do not really know whether we are probing $\rho_{\mu e}$ 
for the BSM benchmark below the strength of $\lambda_e$ yet.
Thus, for example, if $\rho_{tt} = 0$ and 
EWBG is through the $\rho_{tc}$ mechanism~\cite{Fuyuto:2017ewj},
then the MEG bound of Eq.~(\ref{eq:MEG16}) only requires
$\rho_{\mu e} = \rho_{e\mu} \lesssim 1.9 \lambda_e$
for our $h$ benchmark, and MEG~II could probe down to $0.7 \lambda_e$.
Both values  are still in accord with Eq.~(\ref{eq:rhoi1}),
but we note that if $c_\gamma$ is lower than the value of 0.2 used, 
which seems likely, then the allowed $\rho_{\mu e}$ range would rise.

{As a passing remark, we expect $\tau \to e\gamma$ to be
much suppressed compared with $\tau \to \mu\gamma$ in g2HDM,
as $\rho_{\tau e}$ is expected to be much smaller than $\rho_{\tau\mu}$.}

%%%%%%%%%%%%%%%%%%%%%%%%
\section{\boldmath Other $\mu$FV Processes}
%%%%%%%%%%%%%%%%%%%%%%%%

%
\subsection{\boldmath $\mu \to 3e$ and $\tau\to \mu\gamma,\,3\mu$}

As Mu3e would start soon to
finally probe below the old SINDRUM bound of $10^{-12}$, 
Eq.~(\ref{eq:SINDRUM88}), we estimate the $\mu \to3e$ rate.
We find, consistent with Ref.~\cite{Crivellin:2013wna}, 
the simple tree level formula for $\mu \to 3e$, 
{\begin{align}
 {\cal B}(\mu\to 3e)  & = \frac{1}{32}
   \biggl[2\Bigl|\sum\frac{y_{\phi\mu e}^\ast y_{\phi ee}}{ \hat m_\phi^2}\Bigr|^2
          + 2\Bigl|\sum\frac{y_{\phi e\mu}^\ast y_{\phi ee}}{ \hat m_\phi^2}\Bigr|^2 \biggr. \nn\\
	 & \ \ \ \ \biggl. +\, \Bigl|\sum\frac{y_{\phi\mu e}y_{\phi ee}}{\hat m_\phi^2}\Bigr|^2 
       + \Bigl|\sum\frac{y_{\phi e\mu}y_{\phi ee}}{\hat m_\phi^2} \Bigr|^2 \biggr],
\end{align}
where we ignore extra Yukawa coupling corrections to 
the muon decay rate $\Gamma_\mu$~\cite{Hou:2019uxa},
$y_{\phi ij}$ are Yukawa couplings for $\phi = h,\, H,\, A$
that can be read off from Eq.~(3) of Ref.~\cite{Hou:2020tgl},
and $\hat m_\phi$ are scalar masses normalized to $v$.
%

%Although $\mu \to 3e$ is independent of $\rho_{tt}$,
%
In view that 200~GeV may be too aggressive for 
the lowest possible exotic scalar mass,
we take {for illustration} the relatively conservative 
$m_H = m_A = 300$ GeV. 
We define our benchmark further as follows:
we take{, somewhat arbitrarily,} $c_\gamma = 0.05$ for the effect from $h$;
we take $\rho_{\mu e} (= \rho_{e\mu})$, $\rho_{ee}$
and $\rho_{\tau e} (= \rho_{e\tau}) = \lambda_e$ [Eq.~(\ref{eq:rhoi1})], 
and take
$\rho_{\tau\tau}$ and $\rho_{\tau\mu} (= \rho_{\mu\tau})
 = \lambda_\tau$ [Eq.~(\ref{eq:rho3j})].
We then find that $\rho_{tt} \simeq 0.4$ 
saturates the MEG bound on $\mu \to e\gamma$,
{and ${\cal B}(\mu\to3e)|^{\rm contact} \sim 5 \times 10^{-24}$ 
at tree level, which is far out} of experimental reach.
But the $\mu e\gamma$ dipole coupling can generate $\mu \to 3e$~\cite{Kuno:1999jp},
\begin{align}
 {\cal B}(\mu \to 3e) \simeq \frac{\alpha}{3\pi}
    \left[{\rm log}\left(\frac{m_\mu^2}{m_{e}^2}\right) - \frac{11}{4} \right]
 {\cal B}(\mu \to e\gamma),
 \label{eq:mu3e}
\end{align}
and we find ${\cal B}(\mu \to 3e)|^{\rm dipole} \simeq {2.6} \times 10^{-15}$
for our benchmark.
Though out of reach of Mu3e in early phase,
it should be detectable with muon intensity upgrades,
where the experiment should be able to confirm the
$\mu \to e\gamma^* \to 3e$ nature.

For $\tau$, our benchmark gives
{${\cal B}(\tau \to \mu\gamma) \simeq 3.1 \times 10^{-9}$}, which is 
an order of magnitude below current B factory bound,
but reachable by Belle~II.
Using analogous formulas as above,
we find ${\cal B}(\tau \to 3\mu)|^{\rm contact} \simeq 4.9 \times 10^{-13}$,
%for the treel level ``contact'' process.
%Again, the $\tau\mu\gamma$ dipole would dominate, 
and the larger %for our benchmark we obtain 
{${\cal B}(\tau\to3\mu)|^{\rm dipole} \simeq {7.0 \times 10^{-12}}$,} 
which is still out of Belle~II reach.
%Even if the current ${\cal B}(\tau \to \mu\gamma)$ bound is
%saturated, it doesn't seem that 
%
However, if Belle~II discovers $\tau \to \mu\gamma$ in early data, i.e., above $10^{-8}$, 
{which is certainly possible~\cite{Hou:2020tgl}} in g2HDM,
it would imply $\tau\to3\mu$ at $10^{-10}$ or above,
which can be probed by the fixed-target experiment, TauFV~\cite{TauFV}, 
that is being planned.
{Also arising from the $\tau\mu\gamma$ dipole, 
$\tau^- \to \mu^-e^+e^-$ would be slightly higher.
But, suppressed by $\rho_{e\mu}$,} 
the $\tau^- \to \mu^-e^+\mu^-$ process is expected to be 
far below the $\tau \to 3\mu$ contact process in g2HDM, 
while $\tau \to e^-\mu^+\mu^-$ would be suppressed 
by the $\tau \to e\gamma$ dipole transition.

%{We note that$\mu \to e\gamma^* \to 3e$? $\tau \to \mu\gamma^* \to 3\mu$?}
%can be experimentally distinguished.

%
\subsection{\boldmath $\mu N \to eN$ conversion}

With two competing experiments, COMET and Mu2e, prospects 
for pushing $\mu \to e$ conversion during the next decade 
or more is exceptionally bright, as
the current limit~\cite{Bertl:2006up} of  $R_{\mu e} < 7 \times 10^{-13}$, 
Eq.~(\ref{eq:SINDRUM06}), is expected to improve 
by $\sim$ 3--4 orders of magnitude~\cite{Adamov:2018vin,Bartoszek:2014mya}.

The relevant effective Lagrangian is 
given by~\cite{Cirigliano:2009bz,Crivellin:2014cta}
\begin{equation}
	\begin{aligned}
\mathcal{L}_{\mathrm{eff}} =\,
 & m_{\mu} \bigl(C_{T}^{R}\, \bar{e} \sigma_{\alpha\beta} L \mu
                         + C_{T}^{L}\, \bar{e} \sigma_{\alpha\beta} R \mu\bigr) F^{\alpha\beta} \\
% & + \, \bigl(C_{q q}^{V L}\, \bar{e} \gamma_{\alpha} L \mu
%                + C_{q q}^{V R}\, \bar{e} \gamma_{\alpha} R \mu\bigr)\,
%                                                \bar{q} \gamma^{\alpha} q  \\
 & + \, \bigl(C_{q q}^{S R}\, \bar{e} L \mu
                + C_{q q}^{S L}\, \bar{e} R \mu\bigr)\, m_{\mu} m_{q} \bar{q} q, \label{eq:Leff_mue}
\end{aligned}
\end{equation}
where $C_{T}^{L,R}$ correspond to the $\mu e\gamma$ dipole,
while $C_{qq}^{SL(R)}$ are coefficients to contact terms generated by scalar exchange.
There are no current-current interactions at tree level in g2HDM.
One computes the conversion rate $\Gamma_{\mu \to e}$
and normalizes to the muon capture rate to get $R_{\mu e}$. 
The conversion rate is given by 
\begin{equation}
	\begin{aligned}
\Gamma_{\mu \rightarrow e} = {m_{\mu}^{5}}
 & \left| {1\over 2} C_{T}^{L(R)} D +
               2\Bigl[m_{\mu} m_{p}\, \tilde{C}_{p}^{SL(R)} S^{p}
     + p \rightarrow n\Bigr]\right|^{2},
% \\
% & \quad\quad\ + L \rightarrow R.
 \label{eq:Gam_mue}
\end{aligned}
\end{equation}
where the $L$ and $R$ effects add in quadrature, {and
$S^{p,n}$ accounts for lepton-nucleus overlap.}
%, and one averages over the atomic field. 
For gold, we use~\cite{Kitano:2002mt} 
$D = 0.189$, $S^p = 0.0614$, and $S^n=0.0918$.
{In Eq.~(\ref{eq:Gam_mue}),}
\begin{equation}
	\begin{aligned}
%\tilde{C}_{p}^{SL(R)} &= \sum_{q=u, d} C_{q q}^{VL(R)} f_{V_{q}}^{p}, \\
\tilde{C}_{p}^{SL(R)} &= \sum %_{q=u, d, s}
          C_{q q}^{SL(R)} f_{q}^{p},
 \label{eq:Ceff_mue}
\end{aligned}
\end{equation} 
relates to %vector $f_{V_q}^{p,n}$ and 
{nucleon matrix elements, $f_{q}^{p,n}$,}
 that account for the quark content of the proton,
where we use $f_u^p = f_d^n = 0.024$, 
$f_d^p = f_u^n = 0.033$~\cite{Harnik:2012pb}, 
$f_s^p = f_s^n = 0.043$~\cite{Junnarkar:2013ac}. 
For heavy quarks, we follow Ref.~\cite{Harnik:2012pb} and use
$f_Q^{p,n}= (2/27)(1-f_u^{p,n}-f_d^{p,n}-f_s^{p,n})$~\cite{Shifman:1978zn} 
for $Q = c,\,b,\,t$.

In g2HDM, the tree level contribution can be written 
in terms of Wilson coefficients~\cite{Crivellin:2014cta}
for the contact terms induced by the scalar $\phi = h,\, H,\, A$ boson exchange,
\begin{equation}
	\begin{aligned}
		C_{qq}^{SL} &= ({2}/v^4)
         \sum \hat y_{\phi e\mu}{\rm Re}\,\hat y_{\phi qq}/{\hat m_\phi^2},
	\end{aligned}
\end{equation}
where $\hat y_{\phi e\mu}$ ($\hat y_{\phi qq}$) is 
normalized to $\lambda_\mu$ ($\lambda_q$),
and one flips $y_{\phi e\mu} \to y_{\phi \mu e}^\ast$ to get $C_{qq}^{SR}$.
The dipole $C_T^{L,R}$ contributions are related to $\mu\to e \gamma$,
i.e., $C_T^{R,L} = \sqrt{\alpha_e \pi } \,A_{L, R}$, 
where $A_{L, R}$ contribute to ${\cal B}(\mu\to e \gamma$) 
[see Ref.~\cite{Hou:2020tgl} for ${\cal B}(\tau \to \mu\gamma)$ formulas].
%
%Thus, $\mu N \to eN$ conversion can also probe~\cite{deGouvea:2013zba}
%the $\mu \to e\gamma$ dipole transition.
%
The $\mu e\gamma$ dipole again dominates $\mu N \to eN$ conversion,
 with contact terms subdominant.
For our benchmark, we obtain the conversion ratio 
$R_{\mu e}|^{\rm contact} \simeq {2.4 \times 10^{-16}}$ for gold 
as an example, while $R_{\mu e}|^{\rm dipole} \simeq 1.6 \times 10^{-15}$. 
Here, we have used $\rho_{qq} = \lambda_q$ for all quarks,
{except $\rho_{tt} \simeq 0.4$ as inferred from MEG bound with our benchmark}. 
We note that contact terms are relatively important in 
$\mu \to e$ conversion compared to $\mu\to 3e$ process.
These values can be probed at COMET and Mu2e. 
In fact, these experiments are posed to overtake
MEG~II in probing $\mu \to e\gamma$ in g2HDM.
Furthermore, if observed, together with 
{the knowledge of nuclear matrix elements}, 
one can use several different nuclei to probe and 
{extract the effect of the contact term(s) in Eq.~(\ref{eq:Leff_mue})}.

We see that the extra $\rho_{\mu e}$ and 
$\rho_{ee}$ couplings of g2HDM hide very well so far from muon probes. 
It is with the help of extra $\rho_{tt}$ coupling  
via the two loop mechanism~\cite{Chang:1993kw} 
for $\mu \to e\gamma$ decay that MEG 
constrains $\rho_{\mu e} \lesssim \lambda_e$ [see Eq.~(\ref{eq:rhoi1})].
MEG~II would continue this program, 
but the $\mu N \to eN$ experiments, COMET and Mu2e,
would become competitive when $10^{-15}$ sensitivity is reached.
Mu3e can confirm the dipole nature once $\mu \to 3e$ 
is also observed with high muon intensity upgrades.
Likewise, $\tau \to \mu\gamma$ would probe $\rho_{\tau\mu}$ 
modulo $\rho_{tt}$, but the $\tau \to 3\mu$ process seems 
out of reach for Belle~II (hence LHCb) if g2HDM holds,
even if Belle~II quickly observes $\tau \to \mu\gamma$.
Thus, while there remains hope for discovery, $\mu$FV physics look ``sanitized'' 
within g2HDM that possesses these extra $\rho_{\ell\ell'}$ (and $\rho_{tt}$)
Yukawa couplings, which bears witness to the long history of  muon research.

{%\small
%\renewcommand{\arraystretch}{1.3}
%================================================
\begin{table*}[t]
\begin{center}
%================================================
\begin{tabular}{|c|l|l|}
\hline
Decay mode  & \quad\quad Current bound & \quad\quad Future sensitivity   \\	
\hline
\hline
$B_s \to \tau\tau$	&  
\ $5.2\times 10^{-3}$ (LHCb~\cite{Aaij:2017xqt}) 	&  
\ $\sim 8\times 10^{-4}$ (Belle~II,\,$5\,{\rm ab}^{-1}$\,\cite{Kou:2018nap})  	\\
& & \ $\sim 5\times 10^{-4}$ (LHCb~phase~II~\cite{Bediaga:2018lhg}) \\
$B_d\to\tau\tau$	&  
\ $1.6\times 10^{-3}$ (LHCb~\cite{Aaij:2017xqt}) 	&  
\ $\sim 1\times 10^{-4}$ (Belle~II~\cite{Kou:2018nap})  	\\
$B \to K\tau\tau$	& \ $2.3 \times 10^{-3}$ (BaBar~\cite{TheBaBar:2016xwe}) &
\ $\sim 2\times 10^{-5}$ (Belle~II~\cite{Kou:2018nap})  \ \\
\hline
\hline
$B_s \to \tau\mu$	&  
\ $3.4\times 10^{-5}$ (LHCb~\cite{Aaij:2019okb}) 	&  
\ 
{[Not yet publicized]}	\\
$B_d \to \tau\mu$	&  
\ $1.2\times 10^{-5}$ (LHCb~\cite{Aaij:2019okb}) 	&  
\ $1.3 \times 10^{-6}$ (Belle~II~\cite{Kou:2018nap}))  	\\
& & \ $3 \times 10^{-6}$ (LHCb~phase~II~\cite{Bediaga:2018lhg}) \\
$B \to K\tau\mu$	&  
\ $2.8\times 10^{-5}$ (BaBar~\cite{Lees:2012zz}) 	&  
\ $\sim 3 \times 10^{-6}$ (Belle~II~\cite{Kou:2018nap})  	\\
 & \ $3.9 \times 10^{-5}$ (LHCb~\cite{Aaij:2020mqb}) &
\ [LHCb competitive] \\
\hline
\hline
$B_s \to \mu e$	&  
\ $5.4 \times 10^{-9}$ (LHCb~\cite{Aaij:2017cza}) 	&  
\ $3 \times 10^{-10}$ (LHCb~phase~II~\cite{Bediaga:2018lhg})  	\\
$B_d \to \mu e$	&  
\ $1.0 \times 10^{-9}$ (LHCb~\cite{Aaij:2017cza}) 	&  
\ $9 \times 10^{-11}$ (LHCb~phase~II~\cite{Bediaga:2018lhg})  	\\
$B \to K \mu e$	&  
\ $6.4\times 10^{-9}$ (LHCb~\cite{Aaij:2019nmj}) 	&  
\ $\sim 6 \times 10^{-10}$ (LHCb~phase~II~\cite{Bediaga:2018lhg}) \ 	\\
\hline
\hline
$B_s \to \mu\mu$	&  
\ $(3.0\pm 0.4)\times 10^{-9}$ (PDG~\cite{PDG})\; 	&  
\ $ \sim 4.4\%$ (LHCb ($300~{\rm fb}^{-1}$)~\cite{Cerri:2018ypt})  	\\
$B_d \to \mu\mu$	&  
\ $(1.1^{+1.4}_{-1.3})\times 10^{-10}$\,\, (PDG~\cite{PDG}) 	&  
\ $ \sim 9.4\%$ (LHCb ($300~{\rm fb}^{-1}$)~\cite{Cerri:2018ypt}) 	\\
\hline
\hline
$B \to \tau\nu$	&  
\ $(1.1 \pm 0.2)\times 10^{-4}$ (PDG~\cite{PDG}) 	&  
\ $ \sim 5\%$ (Belle~II~\cite{Kou:2018nap})  	\\
$B \to \mu\nu$	&  
\ $(5.3\pm 2.2)\times 10^{-7}$ (Belle~\cite{Prim:2019gtj}) 	&   
\ $ \sim 7\%~({\rm stat})$ (Belle~II~\cite{Kou:2018nap})   	\\
\hline
\hline
\end{tabular}
\caption{
Summary of current experimental data on $B$ decays considered in our analysis. All upper bounds are at 90\% C.L., and phase~II for LHCb stands for
HL-LHC running after upgrade~II.}
\end{center}
\label{tab:B-decays}
\end{table*}
}
%%

%%%%%%%%%%%%%%%%%%%%%%%%
\section{\boldmath Contrast: Muon, or Bold}
%%%%%%%%%%%%%%%%%%%%%%%%

In this section, we contrast the ``sanitized'' muon front of the previous sections
with what we dub the ``bold'' BSM front inspired by $B$ anomalies.
We refer to Ref.~\cite{Hou:2019dgh} for a discussion of
all the current $B$ anomalies, including 
cautionary notes on the experimental results.
Extending from $\mu$FV,
we discuss BSM effects in (semi)leptonic $B$ decays,
be it BSM enhancement of $B_q \to \tau\tau$,
or the purely BSM decays $B_q \to \tau\mu$, $B \to K\tau\mu$.
We also touch upon the $B_q \to \mu\mu$ and $B \to \mu\nu,\, \tau\nu$ decays,
which already appear to be SM-like in rate.

\subsection{\boldmath BSM-enhanced: $B_q \to \tau\tau$ modes}

%The B anomalies have stimulated suggestions of the aforementioned
%decay modes, some of which are quite striking, and have 
%stirred the experiments into action, as it should be.
%
The ``BaBar anomaly'' in $B \to D^{(*)}\tau\nu$~\cite{PDG,Hou:2019dgh}
suggests a large tree level BSM effect interfering with 
the SM $b\to c\tau\nu$ amplitude.
Based on general arguments, it was pointed out~\cite{Capdevila:2017iqn} that 
such a large effect should be accompanied by similar effects in $b \to s\tau\tau$.
Note that, because of the difficult $\tau^+\tau^-$ signature, the experimental
bounds~\cite{PDG} are rather poor.
Projecting from the BaBar anomaly, Ref.~\cite{Capdevila:2017iqn}
suggested that ${\cal B}(B_s \to \tau\tau) \sim 5\times 10^{-4}$ (or larger)
is possible,  to be compared with $\simeq 7.7 \times 10^{-7}$
 in SM~\cite{Bobeth:2013uxa}.
Similarly, ${\cal B}(B \to K^{(*)}\tau\tau) \sim 10^{-4}$ is projected.
The theory suggestion was in part stimulated by the 
LHCb search~\cite{Aaij:2017xqt}, based on 3~fb$^{-1}$ run 1 data, 
setting the 90\% C.L. bound of 
\begin{align}
 {\cal B}(B_s \to \tau\tau) < 5.2 \times 10^{-3}, \quad ({\rm LHCb},\,2017),
 \label{eq:Bstautau}
\end{align}
which is an order of magnitude higher than the theory suggestion. 
Likewise, the only limit on three-body search,
${\cal B}(B^+ \to K^+\tau^+\tau^-) < 2.3 \times 10^{-3}$ 
from BaBar~\cite{TheBaBar:2016xwe}, is also poor.
One suffers from lack of mass reconstruction capability, %in a hadronic environment, 
and only at the HL-LHC after LHCb upgrade~II~\cite{Bediaga:2018lhg} 
can the sensitivity reach $\sim 5 \times 10^{-4}$, 
touching the upper reaches of projected enhancement~\cite{Capdevila:2017iqn}.
Belle~II plans to take some $\Upsilon(5S)$ data early on, and 
projects the reach of  $\sim 8.1 \times 10^{-4}$~\cite{Kou:2018nap}.
As the environment is clean, Belle~II would likely
take more $\Upsilon(5S)$ data if the BaBar anomaly is confirmed.
{For $B \to K^{(*)}\tau\tau$, the Belle~II sensitivity of 
$\sim 2 \times 10^{-5}$~\cite{Kou:2018nap} should be able to 
probe the range of interest at ${\cal O}(10^{-4})$.}

We list the current limits and future prospects for the
$B_q \to \tau\tau$ and $B \to K^{(*)}\tau\tau$ modes in
{Table~II}. %\ref{tab:B-decays}.

\subsection{\boldmath Purely BSM: $B_q \to \tau\mu$ and $B \to K\tau\mu$ modes}

The $B$ anomalies suggest lepton universality violation (LUV),
such as $B \to D^{(*)}\tau\nu$ vs $B \to D^{(*)}\mu\nu$, 
or $B \to K^{(*)}\mu\mu$ vs $B \to K^{(*)}ee$.
It was suggested~\cite{Glashow:2014iga} on general grounds the possibility of 
accompanying lepton flavor violation (LFV), giving rise to interesting decays 
such as $B_q \to \ell\ell'$ and $B \to K\ell\ell'$ for $\ell \neq \ell'$.
As the $B$ anomalies persisted, serious model building went underway,
and we take the so-called PS$^3$ model~\cite{Bordone:2017bld} as 
the standard bearer for ambitious UV-complete models 
(which we term ``bold''). 
To handle severe low energy constraints and focus on the third generation, 
the Pati-Salam (PS) model~\cite{Pati:1974yy} comes in {\it three copies}.
The presence of leptoquarks (LQ) in the Pati-Salam model 
induce the decays such as $B_q \to \tau\mu$ and $B \to K\tau\mu$, 
where detailed phenomenology was given in Ref.~\cite{Bordone:2018nbg}.

These are striking signatures! 
Before long, with 3~fb$^{-1}$ run 1 data, 
LHCb sets~\cite{Aaij:2019okb} the 90\% C.L. limit of 
\begin{align}
 & {\cal B}(B_s \to \tau\mu) < 3.4 \times 10^{-5}, \quad \ ({\rm LHCb},\,2019),
 \label{eq:Bstaumu} %\\
% & {\cal B}(B_d \to \tau\mu) < 1.2 \times 10^{-5}, \quad\quad\quad ({\rm same})
% \label{eq:Bdtaumu}
\end{align}
which contrasts with the poor performance of Eq.~(\ref{eq:Bstautau}) 
for $B_s \to \tau\tau$.
This limit practically ruled out the entire ${\cal B}(B_s \to \tau\mu)$ 
range projected by Ref.~\cite{Bordone:2018nbg}, 
forcing model builders to introduce~\cite{Cornella:2019hct} 
right-handed LQ interaction as tune parameters.
In so doing, $B_s \to \tau\tau$ and $B \to K\tau\tau$ decays 
get enhanced~\cite{Cornella:2019hct}, which is 
in accordance with Ref.~\cite{Capdevila:2017iqn}.
It would be interesting to see the full 9~fb$^{-1}$ run 1\,+\,2 result
for $B_s \to \tau\mu,\, \tau\tau$ modes.
{Perhaps because the analysis of Ref.~\cite{Aaij:2019okb} was still underway 
when the LHCb upgrade~II document~\cite{Bediaga:2018lhg} was being prepared, 
we cannot find the sensitivity projections of $B_s \to \tau\mu$ for full LHCb 
Upgrade~II data (and neither for Belle~II); hence, we state this explicitly in Table~II.}

BaBar has searched~\cite{Lees:2012zz} for the companion $B \to K\tau\mu$ mode. 
Using a full hadronic tag to reconstruct the other charged $B$,
hence with full kinematic control, by measuring $K^+$ and $\mu^-$,
{one projects into the $m_\tau$ window without reconstructing the $\tau$}.
The result at 90\% C.L. is~\cite{Lees:2012zz}
\begin{align}
 {\cal B}(B^+ \to K^+\tau^+\mu^-) & < 2.8 \times 10^{-5}, \ \; ({\rm BaBar},\,2012)
 \label{eq:Kmutau_BB} \\
 &  < 3.9 \times 10^{-5}, \ \;  ({\rm LHCb},\,\, 2020)
 \label{eq:Kmutau_LHC}
\end{align}
for the better measured charge combination, 
and Eq.~(\ref{eq:Kmutau_LHC}) is the recent LHCb 
measurement~\cite{Aaij:2020mqb} with {\it full}\;9\;fb$^{-1}$\,run 1\,+\,2 data.
We first note that Belle has not performed this measurement
so far, despite having more data than BaBar.
The second point to stress is that, although the LHCb result may not appear 
competitive at first sight, they exploit $B_{s2}^{*0} \to B^+K^-$ decay 
and use the $K^-$ to tag~\cite{Stone:2014mza} the $B^+$ for full kinematic control, 
putting LHCb in the game for the $B^+ \to K^+\tau^+\mu^-$ pursuit, 
and making things more interesting for the Belle~II era.

LHCb also places the best bounds~\cite{Aaij:2017cza} for 
${\cal B}(B_s \to \mu e) < 5.4 \times 10^{-9}$ and 
${\cal B}(B_d \to \mu e) < 1.0 \times 10^{-9}$, 
as well as 
${\cal B}(B^+ \to K^+\mu^+e^-) < 6.4 \times 10^{-9}$~\cite{Aaij:2019nmj}.
The current limits and future prospects for the
$B_q \to \tau\mu$ and $B \to K^{(*)}\tau\mu$ modes are listed in
 Table~II. %\ref{tab:B-decays}.
The $\mu e$ counterparts are also listed,
but aside from the comment given in Ref.~\cite{Glashow:2014iga},
it is not easy from the model building point of view
to make projections that are experimentally accessible.

\subsection{\boldmath SM-like: $B_q \to \mu\mu$ and $B \to \tau\nu,\, \mu\nu$ modes}

It is useful to recall that $B_s \to \mu\mu$ 
was a front runner~\cite{PDG} in the 2000's as possibly
greatly enhanced, but a few years into LHC running, 
the $B_{s, d} \to \mu\mu$ decays became consistent with SM:
the PDG values~\cite{PDG} are
 ${\cal B}(B_s \to \mu\mu) = (3.0 \pm 0.4) \times 10^{-9}$ and
 ${\cal B}(B^0 \to \mu\mu) = (1.1^{+1.4}_{-1.3}) \times 10^{-10}$,
compared with the 
{SM expectation~\cite{Beneke:2019slt} of
 ${\cal B}(B_s \to \mu\mu) = (3.66 \pm 0.14) \times 10^{-9}$ and
 ${\cal B}(B^0 \to \mu\mu) = (1.03 \pm 0.05) \times 10^{-10}$.}
We note that ATLAS, CMS, and LHCb have
recently combined~\cite{Amhis} their 2011--2016 data to give
 ${\cal B}(B_s \to \mu\mu) = (2.69^{+0.37}_{-0.35}) \times 10^{-9}$ and
 ${\cal B}(B^0 \to \mu\mu) < 1.6 \times 10^{-10}$ at 90\% C.L.
{A discrepancy for $B_s \to \mu\mu$ at $\sim 2\sigma$ is suggested, 
which was already indicative with PDG average,
while the low value for $B_d \to \mu\mu$ is 
in part due to the negative central value from ATLAS.}
We will use the PDG result (see Table~II), 
which should be good enough for our illustrative purpose.
In any case, the $B_d$ mode is not yet observed, 
but should emerge with sufficient data.
The estimated errors for LHCb at 300 fb$^{-1}$~\cite{Cerri:2018ypt} are given in Table~II. 
Naturally, models such as PS$^3$ do not give large enhancement for $B_q \to \mu\mu$, 
{but $B_s \to \mu\mu$ serves as a reminder of how things might evolve 
for the $B$ anomalies, in as much as these ``anomalies'' are data-driven.}

The $B \to \tau\bar\nu$ rate receives a neat correction~\cite{Hou:1992sy}
in type two 2HDM (2HDM-II), while
%providing a unique constraint on the charged Higgs mass.
Belle measurements~\cite{PDG} have settled around SM expectation, 
and in fact, provides a constraint~\cite{Cornella:2019hct} on PS$^3$.
Since the correction factor of Ref.~\cite{Hou:1992sy}
does not depend on the flavor of the charged lepton, one has the ratio 
$R_B^{\mu/\tau} = {\cal B}(B \to \mu\bar\nu)/{\cal B}(B \to \tau\bar\nu)
\cong 0.0045$ for both SM and 2HDM-II~\cite{Chang:2017wpl}.
But some subtleties 
{such as $V_{tb}/V_{ub}$ enhancement
and the nondetection of neutrino flavor $\bar \nu_i$
(it could be $\bar \nu_\tau$ that escapes)}, as discussed in Ref.~\cite{Hou:2019uxa}, 
allow $R_B^{\mu/\tau}$ to deviate from the expected value precisely in g2HDM, 
and one probes the $\rho_{\tau\mu}\rho_{tu}$ product. 
{Note that our actual knowledge~\cite{Hou:2020ciy} of $\rho_{tu}$ 
is rather poor compared with what is suggested in Eq.~(4).}
The recent Belle update~\cite{Prim:2019gtj} gives
\begin{align}
 {\cal B}(B \to \mu\bar\nu) = (5.3 \pm 2.2) %(5.3 \pm 2.0 \pm 0.9)
      \times 10^{-7}, \ \ ({\rm Belle},\,2020)
 \label{eq:Bmunu}
\end{align}
where we add the statistical and systematic errors in quadrature,
treating as Gaussian.
Equation~(\ref{eq:Bmunu}) is consistent with SM, 
{but gives a two-sided bound, i.e., ${\cal B}(B \to \mu\bar\nu)$ could 
be above or below the nominal SM value~\cite{Hou:2019uxa} of $3.9 \times 10^{-7}$,
%in the range of $\sim (0.3\;{\rm to}\;  1.1) \times 10^{-6}$,
and the $R_B^{\mu/\tau}$ ratio provides a good probe of g2HDM for Belle~II} 
in the next few years.

We reiterate that, though $B_q \to \mu\mu$ are loop processes
while $B \to \tau\nu,\, \mu\nu$ are at tree level,
and the measured values still have to settle,
none are in disagreement with SM expectation,
which put constraints on BSM models inspired by $B$ anomalies, 
as well as g2HDM.
The current status and future prospects are listed in Table~II.
\begin{figure*}[t]
\centering
\includegraphics[angle=0,width=17.8cm]{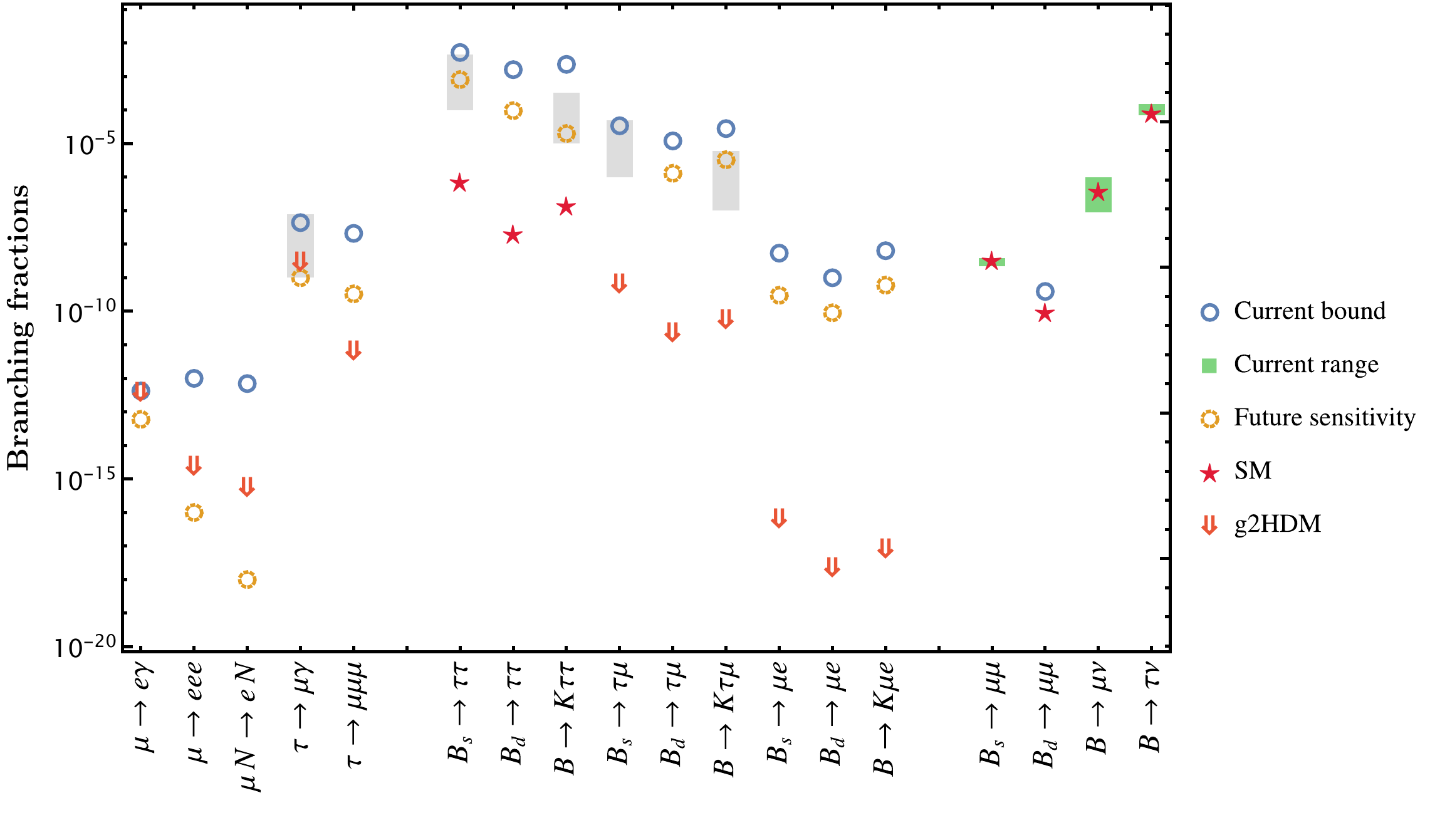}
\caption{
Transcription of Table~II, with
 blue solid circles for current bounds,
 orange dotted circles for future sensitivities,
 green shaded bands for the measured ranges of $B_s \to \mu\mu$ and $B \to \tau\nu,\, \mu\nu$,
 and red $\star$ marking SM predictions.
The grey shaded bands illustrate the five leading predictions of the PS$^3$ model,
while red $\Downarrow$ illustrate g2HDM benchmark projections, where we use 
{$c_\gamma = 0.05$, $m_{H,\, A} = 300$~GeV,  $\rho_{\mu e} = \lambda_e$, 
$\rho_{\tau\mu}=\lambda_\tau$,  and $\rho_{ii} = \lambda_i$}, except $\rho_{tt} = 0.4$. 
%
%The values for $\rho_{sb}$ and $\rho_{db}$ [former is  about ${\cal O}(10^{-3})$, 
%latter is ${\cal O}(10^{-4})$] are taken as allowed from tree-level estimates of 
%$B$-mixings as discussed in. 
See the text for further details.}
\label{contrast}
\end{figure*}

\subsection{Contrasting g2HDM with ``boldness''}

Having presented the status of various (semi)leptonic rare $B$ decays,
where some striking projections arise from models motivated by $B$ anomalies,
we turn to contrasting with g2HDM, the projections of which conform better with 
the more ``sanitized'" tradition of muon physics.

\subsubsection{From $\mu$FV to PS$^3$}

The purely leptonic $\mu$FV processes discussed previously,
such as $\mu \to e\gamma$ in Sec.~II, 
and $\mu \to 3e$, $\tau \to \mu\gamma$, $\tau\to 3\mu$,  
and $\mu N \to eN$ in Sec.~III, are illustrated in Fig.~\ref{contrast}.
That is, the current bounds and future sensitivities listed in Table~I
are plotted as blue solid and orange dotted circles, respectively.
None are so far observed, so the current MEG bound on $\mu \to e\gamma$ 
is also marked by a downward red {\boldmath $\Downarrow$} 
for the g2HDM projection, where, for sake of illustration, we have set up 
a benchmark consistent with Eqs.~(\ref{eq:rho3j}) and (\ref{eq:rhoi1})
and with small $h$-$H$ mixing.
As the scalar-induced contact effect is rather small,
the dipole $\mu \to 3e$ transition is 
also marked by a downward red {\boldmath $\Downarrow$}.
{However, though subdominant, the scalar-induced contact effect 
for $\mu N \to eN$ is not negligible, and 
the downward red {\boldmath  $\Downarrow$} shows
the combined dipole plus contact effect, which is destructive.
The sign of interference, however, could be easily flipped,
so the actual possibilities are considerably broader.}
The $\tau\to \mu\gamma$ rate with this benchmark is also illustrated,
which falls toward the lower range of Belle~II reach,
while we predict that $\tau \to 3\mu$ is out of reach in g2HDM.

Likewise, the current bounds and future sensitivities for 
(semi)leptonic rare $B$ decays discussed in Secs.~III.A and III.B are
also plotted in Fig.~3.
%
%Before we make our comparison, 
%
Of interest here is some {\it two-sided projections},
as they stand at present, for the striking signatures 
arising from PS$^3$~\cite{Cornella:2019hct},
\begin{align}
  10^{-4}  \lesssim {\cal B}(B_s \to \tau\tau) & \lesssim 4.5 \times 10^{-3},
 \label{eq:Bstataps3} \\
  10^{-6}  \lesssim  {\cal B}(B_s \to \tau\mu) & \lesssim 6 \times 10^{-5},
 \label{eq:Bstamups3} \\
  10^{-9}  \lesssim {\cal B}(\tau \to \mu\gamma) & \lesssim 8 \times 10^{-8},
 \label{eq:tamugamps3}
\end{align}
while ${\cal B}(B \to K\tau\mu)$ scales down from 
${\cal B}(B_s \to \tau\mu)$ by a factor of $\sim 9$, 
and for ${\cal B}(B \to K\tau\tau)$ vs ${\cal B}(B_s \to \tau\tau$) 
the factor is $\sim 13$. 
We do not show the ${\cal B}(\tau \to \mu\phi)$ mode~\cite{Cornella:2019hct} 
as it seems out of Belle~II reach. %\gtrsim 3 \times 10^{-12}$.
These ranges are shown in Fig.~\ref{contrast} as grey shaded bands,
where existing bounds for $B_s \to \tau\mu$ and $\tau \to \mu\gamma$
cut into the upper ranges of PS$^3$ projections,
and are the points of our comparison with g2HDM expectations.
{As noted, the future sensitivity for $B_s \to \tau\mu$ is not quite known at present.}

{We note further that, with $\tau \to \mu\gamma$ generated by LQ in the loop,
there is an anticorrelation with ${\cal B}(B_s \to \tau\mu)$
within the PS$^3$ scenario~\cite{Cornella:2019hct}:
if the limit on $B_s \to \tau\mu$ is pushed further down with 9\;fb$^{-1}$
full run 1\,+\,2 data, then ${\cal B}(\tau \to \mu\gamma)$
will move up and become closer to the current limit,
and would be a boon to Belle~II in the model scenario.
Likewise, pushing down on $\tau \to \mu\gamma$
would imply an increased lower bound for $B_s \to \tau\mu,\,\tau\tau$ in PS$^3$.
These bounds and (anti)correlations allow 
the PS$^3$ model to ``provide a smoking-gun signature for this framework \ldots
 or could lead us to rule it out~\cite{Cornella:2019hct}.'' }

{The $B_q \to \mu\mu$ and $B \to \mu\nu,\,\tau\nu$ processes
discussed in Sec.~III.C are plotted differently in Fig.~3, as they are 
now mostly found to be consistent with SM expectations
 (marked as red {\boldmath $\star$}).
The measured $B_s\to\mu\mu$ rate, shown as the narrow green shaded band,
covers the SM expectation but appears slightly on the low side.
Likewise, $B \to \tau\nu$ is also measured to be consistent with SM, 
which Belle~II would continue to probe.
For $B_d \to \mu\mu$, we plot the more conservative upper limit from PDG,
while the latest Belle update on $B \to \mu\nu$ gives a two-sided bound,
which is illustrated by the broad green shaded band that covers the SM expectation.
The PS$^3$ model shies away from processes that involve only muons,
but $B \to \tau\nu$ does provide~\cite{Cornella:2019hct} some constraint.}

%For sake of illustration, we estimate only the tree level contributions
%arising from extra Yukawa couplings of g2HDM
%to 
%

\subsubsection{The $bq\ell\ell'$ processes in g2HDM}

{The rare $B$ decay processes of interest (we only quote results for $B \to \ell\nu$) 
are in the form of $bq\ell\ell'$ four-fermi interactions.}
%We also cover the $b \to d$ processes.
%
Thus, the extra Yukawa couplings that enter on the quark side 
are $\rho_{bs}$, $\rho_{bd}$ at tree level, 
and $\rho_{\ell\ell'}$ for $\ell^{(\prime)} = \tau,\, \mu,\, e$ 
on the charged lepton side.
For the latter, we continue to use our benchmark values
$\rho_{\tau\tau},\, \rho_{\tau\mu} = \lambda_\tau \simeq 0.010$ [Eq.~(\ref{eq:rho3j})], 
and $\rho_{\mu e},\, \rho_{ee} = \lambda_e \cong 0.0000029$ [Eq.~(\ref{eq:rhoi1})]. 
The issue is that, for $\ell = \ell'$, 
 SM loop effects seem affirmed by experiment,
while for $\ell \neq \ell'$, there is no SM loop effect,
and one would need the leptonic FCNH couplings in g2HDM to act.
In the following, we will use tree level approach to $B_q \to \mu\mu$
to infer $B_q \to \ell\ell'$ for $\ell \neq \ell'$ case, while using
loop corrections for $B_q \to \mu\mu$ to discuss $B_q \to \tau\tau$.
{In each case, the corresponding $B_q$ mixing constraints are taken into account.}

{It is well known that the measured~\cite{PDG} $B_q$ mixings can be 
accounted for quite well by SM loop effects}. For example, the operator
$O_1 = (\bar s_\alpha \gamma^\mu L b_\alpha)(\bar s_\beta \gamma_\mu L b_\beta)$
for $B_s$ mixing has coefficient $(G_F m_W V_{ts}^{\ast} V_{tb}/2\pi)^2 S_0(x_t)$,
with $x_t = m_t^2/m_W^2$ and $S_0(x_t)\simeq 2.35$ from SM box diagram,
and one just replaces $s \to d$ for $B_d$ mixing.
{In g2HDM, $\rho_{bq}$ ($q = s,\,d$) enters $B_q$ mixing at {\it tree} level, 
hence stringent constraints are implied.}

%\tcb
{The NP effects in $B_q$ mixings can be parametrized by defining 
$C_{B_q} e^{2 i \Phi_{B_q}}
 = \langle \bar B_q\lvert {\cal H}_{\rm eff}^{\rm Full}\rvert B_q\rangle
  / \langle \bar B_q\lvert {\cal H}_{\rm eff}^{\rm SM}\rvert B_q\rangle$.
Using the 2018 NP fit performed by UTfit~\cite{UTfit2018}, one finds
\begin{equation}
\begin{aligned}
		C_{B_s} &= 1.110 \pm 0.090, \quad
	\Phi_{B_s} = (0.42 \pm 0.89)^{\circ}, \\
	C_{B_d} &= 1.05 \pm 0.11, \quad
	~~~\Phi_{B_d} = (-2.0 \pm 1.8)^{\circ}.
\end{aligned}\label{eq:CBsCBd}
\end{equation}
For sake of illustration 
{and to reduce the number of parameters, 
we will treat extra Yukawas as real}
and assume that adding the g2HDM effect,
$C_{B_q}$ and $\Phi_{B_q}$ stay within 2$\sigma$ ranges
of Eq.~(\ref{eq:CBsCBd}).}

In g2HDM, {the leading effect} %~\cite{Crivellin:2013wna}} 
comes from the operator 
$O_4 = (\bar s_\alpha L b_\alpha)(\bar s_\beta R b_\beta)$ at tree level,
which constrains the product $\rho_{sb}\rho_{bs}^\ast$, 
while the operators 
$O_2 = (\bar s_\alpha L b_\alpha)(\bar s_\beta L b_\beta)$ and 
$O'_2 = (\bar s_\alpha R b_\alpha)(\bar s_\beta R b_\beta)$ 
constrain individual couplings $\rho_{bs}^\ast$, $\rho_{sb}$
but are less constraining.
Furthermore, the coefficients of $O_2^{(\prime)}$ suffer
cancellation between $H$ and $A$ contributions.
{Assuming $O_4$ dominance, one has the coefficient} 
$C_4=  -{y_{\phi b s}^\ast \,y_{\phi s b}}/{m_\phi^2}$, 
 where $\phi$ is summed over $h$, $H$, $A$,
and we take 
{$c_\gamma = 0.05$} and $m_H = m_A = 300$~GeV as before.
Taking renormalization group evolution into account~\cite{Becirevic:2001jj}, 
using bag factors from Ref.~\cite{Carrasco:2013zta}
and decay constants from Ref.~\cite{Aoki:2019cca}, we find
$|\rho_{sb}\rho_{bs}^\ast| \lesssim (0.021 \,\lambda_b)^2$. 
In similar vein, we obtain
$|\rho_{db}\,\rho_{bd}^\ast| \lesssim (0.0046\,\lambda_b)^2$,
where we take $\lambda_b \simeq 0.016$.
Assuming reality, 
{we adopt
 $\rho_{sb} \simeq \rho_{bs}^* \simeq 0.021\lambda_b \sim 0.00034$, and
 $\rho_{db} \simeq \rho_{bd}^* \simeq 0.0046\lambda_b \sim 0.000074$,} 
respectively. 
%The symmetric nature of the $\rho^d$ matrix, 
%analogous to our working assumption for $\rho^\ell$,
%will be justified when we consider $B_s \to \mu\mu$ below.

{With $\rho_{bs}$, $\rho_{bd}$, and $\rho_{\ell\ell'}$ so small, 
one may expect $B_q \to \ell\ell$ modes would be SM-like in g2HDM,
%While $B_q \to \tau\tau$ is still far away from experimental scrutiny,
%we do know that
%
which is the case for $B_s \to \mu\mu$, 
and to some extent $B_d \to \mu\mu$ as well: 
the measured strengths are indeed SM-like.} 
At tree level, we find that $B_s \to \mu\mu$ gives 
stringent constraints on $\rho_{bs(sb)}$, and can be 
on a par with those from $B_s$ mixing constraints. 
For example, {for our benchmark of} 
$c_\gamma =0.05$, $\rho_{\mu\mu}=\lambda_\mu \sim 0.00061$, 
and $m_H=m_A=300$~GeV, the $2\sigma$ range of ${\cal B}(B_s\to\mu\mu)$ 
gives the bound of $\rho_{sb}=\rho_{bs}  \in [-0.019 \lambda_b, 0.143 \lambda_b]~ \vee ~[1.173 \lambda_b, 1.334 \lambda_b] $, which is relaxing than $B_s$ mixing.} 
On the other hand, due to poorer measurement of $B_d \to \mu\mu$ so far, 
bounds on $\rho_{db(bd)}$ from $B_d\to\mu\mu$ are weaker than $B_d$ mixing.
Thus, by the fact that $B_q \to \mu\mu$ rates are already 
SM-like in g2HDM, 
{we expect $B_q \to \tau\tau$ to be 
not so different from SM expectations if tree contributions prevail.}

With $\rho_{sb} = \rho_{bs}$ and $\rho_{db} = \rho_{bd}$ 
so suppressed, one has to take up-type extra Yukawa couplings into account,
which contribute to $B_q$ mixings and $B_q \to \ell\ell$ at one loop order. 
The leading contributions to $B_q$ mixings come from the same box diagrams as SM, 
but with either one $W^+$ or both replaced by $H^+$, which also generates $O_1$. 
Considering the effect of $\rho_{tt}$ only, we obtain 
{$\Delta C_1^{WH} = y x_t V_{ts}^{\ast 2} V_{tb}^2\,|\rho_{tt}|^2
 g(y, y x_t)/32 \pi^2 v^2$, where $y = M_W^2/m_{H}^2$} 
for the $WH$ box correction, and 
$\Delta C_1^{HH}= -V_{ts}^{\ast 2}V_{tb}^2 \,|\rho_{tt}|^4
 f(y x_t)/128 \pi^2 m_{H}^2$ for the $HH$ box correction. 
{Here, $H$ stands as shorthand for $H^+$},
and the loop functions $f$ and $g$ are given in the Appendix. 

Considering this one loop contribution by itself
gives a constraint on the $\rho_{tt}$--$m_{H^+}$ plane. 
For example, for a $300$\;GeV charged Higgs boson, 
we find $|\rho_{tt}|\lesssim 0.8$, and similar bound from $B_d$ mixing as well. 
However, we caution that inclusion of additional up-type {Yukawa couplings} 
can induce cancellation effects, thereby weakening the constraint. 
Most notably, with $\rho_{ct}$ as small as ${\cal O}(10^{-2})$, 
one can relax $\rho_{tt}$ to $\sim 1 $. 
As stated, we avoid cancellations and discuss tree and loop contributions separately. 
The same treatment is applied to rare $B$ decays,
and we continue to assume $\rho_{qb} = \rho_{bq}$ and take them as real.

$B_q\to\mu\mu$ can also receive significant contribution 
through one loop diagrams, where the leading effect is from 
$Z$ penguins with $H^+$ and top in the loop. 
This is a lepton flavor universal contribution and modifies the coefficient of
 $O_{10} = (\bar s \gamma^\alpha L b) (\bar \ell \gamma_\alpha \gamma_5 \ell)$. 
We find~\cite{Crivellin:2019dun} the $\rho_{tt}$ correction 
{$\Delta C_{10}^{H^+} = |\rho_{tt}|^2 h(y x_t)/16 \pi \alpha_e$}, 
where the loop function $h$ is given in the Appendix. 
The other loop diagrams 
{are suppressed in the small $\rho^d_{ij}$ approximation 
and/or by extra lepton $\rho^\ell$ Yukawa couplings (such as in box diagrams)}. 
Similar to $B_q$ mixing, $\Delta C_{10}^{H^+}$ puts a
constraint on the $\rho_{tt}$--$m_H^+$ plane. 
For $m_H=m_A=300$ GeV, we obtain $\rho_{tt} \lesssim 0.4$ 
for $2\sigma$ range of  ${\cal B}(B_s\to\mu\mu)$,
which is 
{more stringent than $B_{s}$ mixing}. 
However, as already noted, the bound weakens 
if one includes other extra Yukawa couplings such as $\rho_{ct}$, 
which receives $|V_{cs}/V_{ts}|$ enhancement. 
In our numerical analysis, we therefore keep the tree level and one loop 
discussions separate, and only comment on cancellation effects later. 
Since LFV decays such as $B_s \to \ell\ell^\prime$ for $\ell\ne \ell^\prime$ 
arise at tree level in g2HDM, we give 
{tree level upper reaches with $\rho_{sb}$ and $\rho_{bs}$ 
satisfying $2\sigma$ range of $B_s$ mixing and $B_s\to\mu\mu$. } 
%As the formalism is a little more involved, we show in the Appendix
%that the tree level corrections to $B_s \to \mu\mu$ are proportional to
%$\rho_{sb} - \rho_{bs}^*$ in amplitude, so would vanish in the
%real symmetric limit. Taking this ansatz, one would find
%$B_d \to \mu\mu$ and $B_q \to \tau\tau$ too all be SM-like.

%We therefore focus on the flavor violating $B_q \to \tau\mu$ 
%and $B \to K\tau\mu$ decays in the following.

%
The effective Hamiltonian  {for flavor violating $B_s$} $\to \tau\mu$ 
and $B \to K\tau\mu$ decays is of the form~\cite{Becirevic:2016zri},
\begin{equation}
	{\cal H} = - (C_S O_S + C_P O_P + C'_S O'_S + C'_P O'_P),
\end{equation}
where
\begin{align}
 \mathcal{O}_{S} & = (\bar s R b)(\bar\ell \ell'),  \quad\;\
 \mathcal{O}_{P}     = (\bar s R b)(\bar\ell \gamma_{5} \ell'),
 \label{eq:op_bsll'}
\end{align}
and ${\cal O}'_{S,P}$ are obtained by exchanging $L\leftrightarrow R$. 
Although $C_S$ and $C_P$ vanish for $\ell = \ell'$ in SM,
tree level exchange of scalar bosons in the g2HDM lead to
\begin{eqnarray}
	C_{S,P}^{\ell\ell'}
 &=& \sum {y_{\phi sb}
             (y_{\phi \ell\ell'} \pm y_{\phi \ell' \ell})}/{2m_\phi^2}, %\\
%	C_S^{\prime\,\ell\ell^\prime} &=& \frac{\sqrt{2}}{8 G_F V_{ts}^\ast V_{tb}}\sum_\phi %\frac{y_{\phi sb}^\ast(y_{\phi \ell\ell^\prime}+y_{\phi \ell^\prime \ell})}{m_\phi^2},\\
%	C_P^{\prime\,\ell\ell^\prime} &=& \frac{\sqrt{2}}{8 G_F V_{ts}^\ast V_{tb}}\sum_\phi %\frac{y_{\phi sb}^\ast(y_{\phi \ell\ell^\prime}-y_{\phi \ell^\prime \ell})}{m_\phi^2}.
 \label{eq:CSll'}
\end{eqnarray}
with $\phi$ summed over $h$, $H$ and $A$, 
and $C_{S,P}^{\prime\,\ell\ell'}$ is obtained from $C_{S,P}^{\ell\ell'}$
by changing $y_{\phi sb} \to y_{\phi bs}^\ast$.

For $B_s\to \ell\ell'$ decay, we use~\cite{Becirevic:2016zri}
\begin{align}
 & \mathcal{B}(B_{s}\rightarrow \ell \ell') \simeq
     \frac{f_{B_s}^2 m_{B_s}  \lambda^{1/2}(m_{B_s}, m_\ell, m_{\ell'})}
                {32 \pi (m_b + m_s)^2\,\Gamma_{B_s}^{\rm heavy}}  \nn \\
 & \times \left[(m_{B_s}^2 - m_+^2)|\Delta C_S|^2
                   +  (m_{B_s}^2 - m_-^2)|\Delta C_P|^2\right],
\end{align}
where $\lambda(a, b, c) = [a^2-(b-c)^2][a^2-(b+c)^2]$, 
{$\Gamma_{B_s}^{\rm heavy}$ is the decay width of the heavy $B_s$ state},
$m_\pm = m_\ell\, \pm\, m_{\ell^\prime}$,  and  $\Delta C_i = C_i - C_i^\prime$. 
With our benchmark of $c_\gamma = 0.05$, $m_H = m_A = 300$~GeV, 
 and leptonic couplings, and the allowed range of $\rho_{sb, bs}$ 
extracted from flavor conserving $B_q\to \mu\mu$ 
(and in conjunction with bounds from $B_s$ mixing), 
{the projections} of various LFV $B$ decays in g2HDM are 
given in Fig.~\ref{contrast} as red {\boldmath $\Downarrow$}. 
%
%In any case, inspection of Eq.~(\ref{eq:CSll'}) reveals
%that the same ansatz of $\rho_{sb} = \rho_{bs}^*$
%that assures $B_s \to \mu\mu$ would be SM-like,
%would lead to very small ${\cal B}(B_s\to \tau\mu)$,
%which vanishes as $|\rho_{sb} - \rho_{bs}^*|^2$.
%
%Even for $|\rho_{sb} - \rho_{bs}^*| \simeq \rho_{sb} \sim 0.12\lambda_b$,
%the largest ${\cal B}(B_s\to \tau\mu)$ one gets is ${\cal O}(10^{-7}$).
%
Analogously, for $B\to K \ell\ell'$, we use~\cite{Becirevic:2016zri}
\begin{equation}
	\begin{aligned}
 & {{d} \mathcal{B}}({B} \rightarrow {K} \ell \ell')/{{d} q^{2}}
   =  \mathcal{N}_{K}^{2} \sum_{i = S, P} \varphi_{i}\, |C_{i}+C'_{i}|^{2},
\end{aligned}
\end{equation}
where $\varphi_S$ is a function of $B\to K$ form factors 
and $\mathcal{N}_K$ a normalization factor. 
Both are $q^2$ dependent, and explicit expressions 
can be found in Ref.~\cite{Becirevic:2016zri}.
%The same remark applies for replacing $C_S$ by $C_P$ 
%if $\rho_{\ell\ell'} = -\rho_{\ell'\ell}$ instead of $\rho_{\ell\ell'} = \rho_{\ell'\ell}$.

\subsubsection{Comparing g2HDM with PS$^3$}

Let us now make the comparison of 
{the spectacular PS$^3$ projections  with the modesty of g2HDM.}
%between g2HDM and PS$^3$.
%
%As we have taken the ansatz of
%real symmetric $\rho^d$ Yukawa matrix to keep
%$B_s \to \mu\mu$ around SM expectation
% (i.e. change of ${\cal B}(B_s \to \mu\mu)$ from SM
% would vanish with $\rho_{sb} - \rho_{bs} \to 0$),

%
We have taken a simplified approach of
%As we have shown that flavor violating couplings $\rho_{sb}$ and $\rho_{bs}$ 
%are very tiny from considerations of 
treating $B_s\to\mu\mu$ and $B_s$ mixing either at tree level,
or at one loop level, but not both simultaneously. 
Either way, the fact that
$B_s \to \mu\mu$ is already consistent with SM expectation
implies $B_s \to \tau\tau$ in g2HDM should also be SM-like{,
which is more so if loop is dominant}. 
This is in contrast with the sizable enhancement projected in 
PS$^3$ (grey shaded band in Fig.~\ref{contrast}),
which can be probed by LHCb upgrade~II,
or dedicated runs by Belle~II on $\Upsilon(5S)$.
For g2HDM, some enhancement (or suppression)
of $B_s \to \tau\tau$ is possible, given that tree effect is controlled
by {$\rho_{\tau\tau}$} which is at ${\cal O}(\lambda_\tau)$,
while tree effect for $B_s \to \mu\mu$ is controlled
by {$\rho_{\mu\mu}$} which is at ${\cal O}(\lambda_\mu)$.
{But these order of magnitude estimates suggest that
bridging the 2 orders of magnitude gap is unlikely},
and g2HDM should be distinguishable from PS$^3$.
In any case, {measurement of $B_s\to \tau\tau$ is a challenge},
while prospects for $B_d \to \tau\tau$ at Belle~II remains to be seen.

More promising for PS$^3$-type of models would be
$B_s \to \tau\mu$, which can saturate the current bound, 
and the discovery, 
{perhaps even with run $1\,+\,2$ data of LHCb,
 would be truly spectacular.}
Projections for g2HDM, however, appears quite out of reach,
as it is 3 orders of magnitude below the lower reach of the PS$^3$ projection.
But our previous caution applies, that an order of magnitude
enhancement is 
not impossible, though it would still be far out of reach. 
In addition, if one allows cancellation
between tree and loop effects in both $B_s \to \mu\mu$
and $B_s$ mixing, it is not impossible that $\rho_{bs(sb)}$ 
can be larger than our suggested values,
{resulting in possible further enhancement of $B_s \to \tau\mu$}.
The challenge is with experiment.
As we noted in Table~II, the projected sensitivities,
be it for LHCb or Belle, are 
{not known publicly.}

At this point, we remind the reader of the ``seesaw'' between
$B_s \to \tau\mu$ and $\tau \to \mu \gamma$ within PS$^3$~\cite{Cornella:2019hct}.
Depending on analysis prowess and/or data accumulation speed,
either measurement could be improved substantially in the next couple of years.
{If one limit is pushed down, %likely on $B_s \to \tau\mu$ by LHCb, 
then the prospect for the other would rise in PS$^3$.} 
In contrast, for g2HDM, while there is 
discovery potential for $\tau\to \mu\gamma$, 
one does not expect $B_s \to \tau\mu$ to be observed any time soon.
The situation for the $B \to K\tau\mu$ mode is similar,
%that g2HDM projections fall below PS$^3$ and can
%vanish with symmetric $\rho^d$ Yukawa matrix.
where the projected sensitivity is again not yet clear,
and we have given the number for Belle~II in Table~II,
which barely starts to touch the PS$^3$ range.
The situation for $B_d \to \tau\mu$ in g2HDM would correlate
with the outcome of $B_d \to \mu\mu$ measurement, 
while the PS$^3$ model does not provide predictions.
Neither models foresee $B_q \to \mu e$ and $B \to K\mu e$ modes 
to be observable. 
{Our projections for g2HDM are given in Fig.~\ref{contrast}.}

As we have also listed in Fig~\ref{contrast}, 
$B \to \mu\bar\nu$ provides a unique probe~\cite{Hou:2019uxa} of g2HDM,
while $B \to \tau\bar\nu$ again appears SM-like already.
These are charged $B$ decays, in contrast to neutral $B$ decays for $B_q \to \ell\ell'$.
As a reminder for purely leptonic $\mu$FV processes,
the $\mu\to e\gamma$, $\mu N \to eN$ and $\tau \to \mu\gamma$
processes have discovery potential, all basically probing the $\mu e\gamma$
{and $\tau\mu\gamma$} dipoles in g2HDM,
{though the $\mu N \to eN$ process can pick up contact effects}. 
In contrast, $\mu \to 3e$ and $\tau \to 3\mu$ would be higher order effects
of the respective dipole transitions.
We mention in passing that muon $g-2$ would not be affected in g2HDM,
while muon EDM, $d_\mu$, would likely scale by $m_\mu/m_e \sim 200$,
and $|d_\mu| \lesssim 2 \times 10^{-27}\; e\,$cm seems, 
{unlike electron EDM $d_e$, far out of experimental reach}.

%%%%%%%%%%%%%%%%%%%%%%%
\section{Discussion and Conclusion}
%%%%%%%%%%%%%%%%%%%%%%%

There are good reasons to take g2HDM, the general two Higgs doublet model
with extra Yukawa couplings, very seriously.
By discovering the $h$ boson and finding that it
closely resembles the SM Higgs boson,
%together with the already known three Goldstone boson components
%of the observed massive vector gauge bosons, 
we now have one weak scalar doublet.
Whether by Gell-Mann's totalitarian principle~\cite{Gell-Mann:1956iqa} 
or the principle of plentitude~\cite{Tot},
with the existence of one scalar doublet, 
there {\it should} be a second doublet, and 
by the same argument, extra Yukawa couplings.
To declare~\cite{Glashow:1976nt} {\it natural} flavor conservation (NFC)
and forbid extra Yukawa couplings, or using a $Z_2$ symmetry to implement it,
are not only {\it not natural} but quite {\it ad hoc} or artificial.
Had supersymmetry (SUSY) emerged at the LHC, 
it would have given credence to 2HDM-II, a type of 2HDM 
with $Z_2$ symmetry to forbid extra Yukawa couplings. 
But the lack of evidence for SUSY so far~\cite{PDG}
suggests that the SUSY scale is considerably above $v$, 
the electroweak symmetry breaking scale.

With three types of charged fermions, each coming in three generations, 
and that the extra Yukawa couplings are naturally complex, 
one has {\it 54 new Yukawa couplings}, which may appear excessive.
There are also seven new Higgs parameters, which include
the $h$-$H$ mixing parameter $c_\gamma$,
and the exotic Higgs masses $m_H$, $m_A$, and $m_{H^+}$.
But the increment of 54 new flavor parameters is on top of 
the existing plentitude of 13 within SM, while the {\it structure} built-in
by nature %into these parameters 
seems to have helped ``obscure'' the presence of the extra Higgs sector parameters:
as we have stated, $m_H$, $m_A$ and $m_{H^+}$ in g2HDM
{\it naturally} populate the 300--600~GeV range. The latter follows
if one takes~\cite{Hou:2017hiw} the principle that 
all dimensionless parameters in the Higgs potential are ${\cal O}(1)$ in strength,
with $v$ as the only scale parameter.
It is curious to note that, with $\rho_{tt}$ naturally ${\cal O}(1)$
because it is a cousin to $\lambda_t \cong 1$,
it may help keep $c_\gamma$ small~\cite{Hou:2017vvp}. 
So the alignment phenomenon may be emergent,
%It is also important to recall that
%, together with ${\cal O}(1)$ Higgs sector parameters, 
while $\rho_{tt}$ could drive EWBG quite effectively.
At any rate, and as we have emphasized,
the flavor parameter structure seems to have hidden itself rather well
from our view, obscuring also the extra Higgs bosons,
which we know so little about.

The flavor structure was first revealed in the 1970s 
through the fermion mass hierarchy, %for all types of charged fermions,
although the existence of three generations triggered Ref.~\cite{Glashow:1976nt}.
But then the mixing hierarchy of $|V_{ub}|^2 \ll |V_{cb}|^2 \ll |V_{us}|^2$
came as a surprise in the early 1980s,
which led to the Cheng-Sher ansatz~\cite{Cheng:1987rs},
suggesting that NFC may be too strong an assumption.
Unknown back then was nature's further design of alignment,
which suppressed FCNH coupling effects of the light, SM-like $h$ boson.
As we stressed in the Introduction, at this point one may find 
fault in the near diagonal nature of the $\rho^d$ Yukawa matrix: 
Why would nature turn off the FCNH effects 
precisely in the sector that we have the best access to?
It is a mystery. But nature has her mysterious ways,
and as an experimental science we can only probe further.
%
%In this context, we mention in passing that we have not covered
%kaon LFV decays, as Nature has shown her method of control.

In summary, 
the extra Yukawa couplings of g2HDM has the built-in 
mass-mixing hierarchy protection as exemplified by Eqs.~(4) and (6), 
plus near diagonal $\rho^d$ Yukawa matrix and alignment.
The $\mu \to e\gamma$ and $\tau \to \mu \gamma$ processes probe
$\rho_{\mu e}\rho_{tt}$ and  $\rho_{\tau \mu}\rho_{tt}$
via the two loop mechanism, and generate 
$\mu \to 3e$ and $\tau \to 3\mu$ at higher order.
The $\mu N \to eN$ process probes the combined effect of 
dipole plus contact terms, and by nature of the process and
experimental prowess, one might disentangle the two effects.
As a second theme, we do not expect LUV or LFV effects to be 
observed soon in (semi)leptonic rare $B$ decays for g2HDM.
This is in contrast with the UV-complete PS$^3$ model
that is the epitome of the recent $B$ anomalies,
where the modes to watch are $B_s \to \tau\mu$, $B \to K\tau\mu$,
and to a lesser extent, $B_s \to \tau\tau$, $B \to K\tau\tau$;
discovering only $\tau \to \mu\gamma$ does 
not distinguish between the two scenarios.
For g2HDM, besides the aforementioned $\mu$FV processes,
$B \to \mu\nu$ may be the mode to watch,
which probes $\rho_{\tau\mu}\rho_{tu}$.

\vskip0.2cm
\noindent{\bf Acknowledgments} \
We thank Jack Chen, Gino Isidori, Matt Rudolph,  
and Sheldon Stone for communications.
This research is supported by 
MOST 106-2112-M-002-015-MY3, 108-2811-M-002-626 of Taiwan, 
and NTU~109L104019.

% \vskip1.6cm

\appendix

\section{Loop functions}
The loop functions for $B_q$ mixing and $B_q \to \ell\ell$ are~\cite{Crivellin:2019dun}
\begin{eqnarray}
		f(a) &=&- \frac{1+a}{(a-1)^2} + \frac{2a\log{a}}{(a-1)^3}, \\
		g(a, b) &=& \frac{1}{(a-b)^2}\left[-\frac{3 a^2 \log (a)}{a-1}
                          + \frac{(b-4 a) (b-a)}{b-1}\right.\nn \\
		           && + \left.\frac{\left(-4 a^2+3 a b^2+2 a b-b^2\right) \log (b)}{(b-1)^2}\right], \\
        h(a) &=& \frac{-a}{a-1} + \frac{a\log a}{(a-1)^2}.
\end{eqnarray}
%

%-----------------------------------------------------------------------------------------------------------------------------------

\end{document}